%% file: main-ELS.tex
\documentclass[final,5p,times,twocolumn]{elsarticle}

\usepackage{amssymb}

\usepackage{xcolor}
\definecolor{main-color}{rgb}{0.6627, 0.7176, 0.7764}
\RequirePackage[english]{babel}   
\usepackage{array, booktabs, makecell}
\usepackage{multirow}
\usepackage{listings}
\usepackage[super]{nth}
\usepackage{verbatim} 

\usepackage[nolist]{acronym}  

\usepackage{tikz,tikzscale}
\usepackage{pgfplots,lipsum}
\usepgfplotslibrary{statistics}

\pgfplotsset{
  compat = 1.9,  
  width=\textwidth,
  every tick label/.append style={
    font=\footnotesize,
    align=center,
  },
  every axis/.append style={
    label style={font=\footnotesize, inner sep=0pt, outer sep=0pt},
  },
  legend style = {text=black,font=\small},
}

\RequirePackage{tum/tum_colors}
\colorlet{blue}{sec_blue_dark}
\colorlet{green}{emph_green}
\colorlet{red}{emph_orange}
\colorlet{orange}{emph_orange}

\definecolor{TUMBeamerDarkRed}   {rgb}{0.79,0.13,0.25}
\usepackage[ruled,vlined]{algorithm2e}

\newcommand{\quotes}[1]{``#1''}

\newcommand{\good}[1]{\textcolor{green}{\textbf{#1}}}
\newcommand{\bad}[1]{\textcolor{red}{\textbf{#1}}}

\usepackage{mathtools}
\usepackage[binary-units]{siunitx}

\usepackage{tikz}
\usetikzlibrary{positioning, shapes.multipart}
\usepackage{bbding} 

\usepackage{blindtext}

\usepackage{subcaption}

\usepackage[export]{adjustbox}

\usepackage{hyperref}
\usepackage{cleveref} 
\crefname{section}{§}{§§}
\Crefname{section}{§}{§§}

\begin{document}

\begin{frontmatter}




\title{REACT: Autonomous Intrusion Response System for Intelligent Vehicles}

\author [a]{Mohammad Hamad \corref{cor1}}
\ead{mohammad.hamad@tum.de}
\author [a]{Andreas Finkenzeller}
\ead{andreas.finkenzeller@tum.de}
\author [a]{Michael K\"uhr}
\ead{michael.kuehr@tum.de}
\author [b]{Andrew Roberts}
\ead{andrew.Roberts@taltech.ee}
\author [c]{Olaf Maennel}
\ead{olaf.maennel@adelaide.edu.au}
\author[d]{Vassilis Prevelakis}
\ead{prevelakis@ida.ing.tu-bs.de}
\author[a]{Sebastian Steinhorst}
\ead{sebastian.steinhorst@tum.de}
 \cortext[cor1]{Corresponding author.}
\affiliation[a]{organization={Technical University of Munich},
            city={Munich},
            country={Germany}}
\affiliation[d]{organization={Technical University of Braunschweig},
        	city={Braunschweig},
        	country={Germany}}
\affiliation[b]{organization={Tallinn University of Technology},
        	city={Tallinn},
        	country={Estonia}}
\affiliation[c]{organization={University of Adelaide},
        	city={Adelaide},
        	country={Australia}}

\begin{abstract}

Autonomous and connected vehicles are rapidly evolving, integrating numerous technologies and software. This progress, however, has made them appealing targets for cybersecurity attacks. As the risk of cyber threats escalates with this advancement, the focus is shifting from solely preventing these attacks to also mitigating their impact.  Current solutions rely on vehicle security operation centers, where attack information is analyzed before deciding on a response strategy. However, this process can be time-consuming and faces scalability challenges, along with other issues stemming from vehicle connectivity. This paper proposes a dynamic intrusion response system integrated within the vehicle. This system enables the vehicle to respond to a variety of incidents almost instantly, thereby reducing the need for interaction with the vehicle security operation center. The system offers a comprehensive list of potential responses, a methodology for response evaluation, and  various response selection methods.  The proposed solution was implemented on an embedded platform. Two distinct cyberattack use cases served as the basis for evaluating the system. The evaluation highlights the system’s adaptability, its ability to respond swiftly, its minimal memory footprint, and its capacity for dynamic system parameter adjustments. The proposed solution underscores the necessity and feasibility of incorporating dynamic response mechanisms in smart vehicles. This is a crucial factor in ensuring the safety and resilience of future smart mobility.

\end{abstract}




\end{frontmatter}



{

	\input{sections/acronyms}
	\input{sections/intro.tex}

	\input{sections/frm}

\input{sections/responseevalaution.tex}
	\input{sections/selection.tex}

	\input{sections/sysarch.tex}
	
	\input{sections/eval.tex}

	\input{sections/conclusion}

	\section*{Acknowledgment}
	 This work is supported by the European Union-funded projects CyberSecDome (Agreement No.: 101120779). 
}





\bibliographystyle{elsarticle-harv} 
\bibliography{sample-base}  





\section*{Biographical Sketches}
\paragraph{\textbf{Mohammad Hamad}}  He has been a research group leader with the Embedded Systems and Internet of Things Group at the Faculty of Computer Engineering, Technical University of Munich, Munich, Germany since 2020. He received his B.Eng. degree in Software Engineering and Information Systems from Aleppo University, Aleppo, Syria, in 2009. He also earned his Ph.D. (Dr.-Ing.) degree in Computer Engineering from the Institute for Data Technology and Communication Networks, Technical University of Braunschweig, Braunschweig, Germany, in 2020. His research interests lie in the area of autonomous vehicles and IoT security.

\paragraph{\textbf{Andreas Finkenzeller}} He received the B.Sc. and M.Sc. degrees in electrical engineering and computer science from Technical University Munich, Munich, Germany, in 2018 and 2021, respectively, where he is currently pursuing the Ph.D. degree with the Embedded Systems and Internet of Things Group.
His research interests include embedded systems, secure communication, and IoT Security.

\paragraph{\textbf{Michael K\"uhr}} He received a B.Eng. degree in Electrical Engineering from the Baden-Wuerttemberg Cooperative State University in Stuttgart, Germany, in 2017 and a M.Sc. degree in Electrical Engineering and Information Technology from the Technical University of Munich, Munich, Germany, in 2022. His research interest focuses on the development and security of automated vehicles.

\paragraph{\textbf{Andrew Roberts}}
He  received the MCyberSecOps from University of New South Wales, Canberra, Australia in 2018 and the MSc degree in cybersecurity engineering from Tallinn University of Technology in 2020. He is currently pursuing the Ph.D. degree in information technology with the Tallinn University of Technology, Estonia. His current research is focussed on cybersecurity testing approaches to autonomous driving algorithms and methods to improve robustness of the design of autonomous systems to cyber threats. 

\paragraph{\textbf{Olaf Maenne}} 
He got his PhD from the Technical University in Munich, studying wide-area Computer Networks and Network security through active and passive measurements and large-scale experiments.  He has since then held faculty positions at Loughborough University in England and Tallinn University of Technology (TalTech) in Estonia, where he led the research at the Centre for Digital Forensics and Cybersecurity and established a Centre for Maritime Cybersecurity in Estonia. Since 2023, he has been with the University of Adelaide. His research interests have broadened over the years to include cyber defence technical exercises and critical infrastructure protection. He has been chairing numerous conferences, including ACM SIGCOMM in London in 2015 and the ACM Internet Measurement Conference (2017), and he is treasurer at ACM SIGCOMM 2024 in Sydney.
\paragraph{\textbf{Vassilis Prevelakis}} 
He received the B.Sc. degree (Hons.) in mathematics and computer science and the M.Sc. degree in computer science from the University of Kent, Canterbury, U.K., in 1984 and 1986, respectively, and the Ph.D. degree in computer science from the University of Geneva, Geneva, Switzerland, in 1996.
He has worked in various areas of security in Systems and Networks both in his current academic capacity and as a freelance consultant. He is the Professor of Embedded Computer Security with the Technical University of Braunschweig, Braunschweig, Germany. His current research involves issues related to vehicular automation security, secure processors, security aspects of software engineering, and auto-configuration issues in secure VPNs.
\paragraph{\textbf{Sebastian Steinhorst}} 
He received the M.Sc. (Dipl.-Inf.) and Ph.D. (Dr. phil. nat.) degrees in computer science from Goethe University Frankfurt, Frankfurt, Germany, in 2005 and 2011, respectively.
He is an Associate Professor with the Technical University of Munich, Munich, Germany, where he leads the Embedded Systems and Internet of Things Group, Department of Electrical and Computer Engineering. He was also a Co-Program PI in the Electrification Suite and Test Lab of the research center TUMCREATE in Singapore. The research of Prof. Steinhorst centers around design methodology and hardware/software architecture co-design of secure distributed embedded systems for use in IoT, automotive and smart energy applications.

\section*{Author Contributions}
Mohammad Hamad: Conceptualization, Methodology, Writing – original draft, Funding acquisition, Supervision. 
Andreas Finkenzeller:  Methodology, Writing – review \& editing. 
Michael K\"uhr: Conceptualization, Methodology, Investigation, Software. 
Andrew Roberts: Validation, Writing – review \& editing. 
Olaf Maenne: Validation, Writing – review \& editing. 
Vassilis Prevelakis: Methodology, Validation, review \& editing.
Sebastian Steinhorst: Funding acquisition, Supervision, Writing – review \& editing.

\end{document}

%% file: sections/acronyms.tex
\begin{acronym}
\acro{ECU}{Electronic Control Unit}
\acro{IRS}{Intrusion Response System}
\acro{IDS}{Intrusion Detection System}
\acro{SAW}{Simple Additive Weighting}
\acro{SOC}{Security Operations Center}
\acro{VSOC}{Vehicle Security Operation Center} 
\acro{E/E}{Electric and Electronic}
\acro{ADAS}{Advanced Driver Assistance Systems}
\acro{TARA}{Threat Analysis and Risk Assessment}
\acro{V2V}{Vehicle-to-Vehicle}
\acro{V2X}{Vehicle-to-Everything}
\end{acronym}

%% file: sections/intro.tex
\section{Introduction}

In recent years, there has been remarkable progress in the development of smart vehicles. Today's vehicles resemble interconnected networks on wheels, with numerous embedded computers, called \acp{ECU}, linked through various types of networks, hosting an extensive number of software components totaling over a hundred million lines of code. Moreover, these networks incorporate various intelligent sensors (such as cameras, LiDAR, radar, etc.) and different connectivity technologies that enhance the vehicle's ability to perceive and interact with the surrounding environment, thus bolstering autonomy and minimizing the reliance on human intervention.
However, with the rise of connectivity and the softwarization of vehicles, the vulnerability to cyberattacks targeting these systems has also escalated \cite{upstream2022}.

 \begin{figure*}[t!]
 	\centering
 	\includegraphics[width=0.99\textwidth]{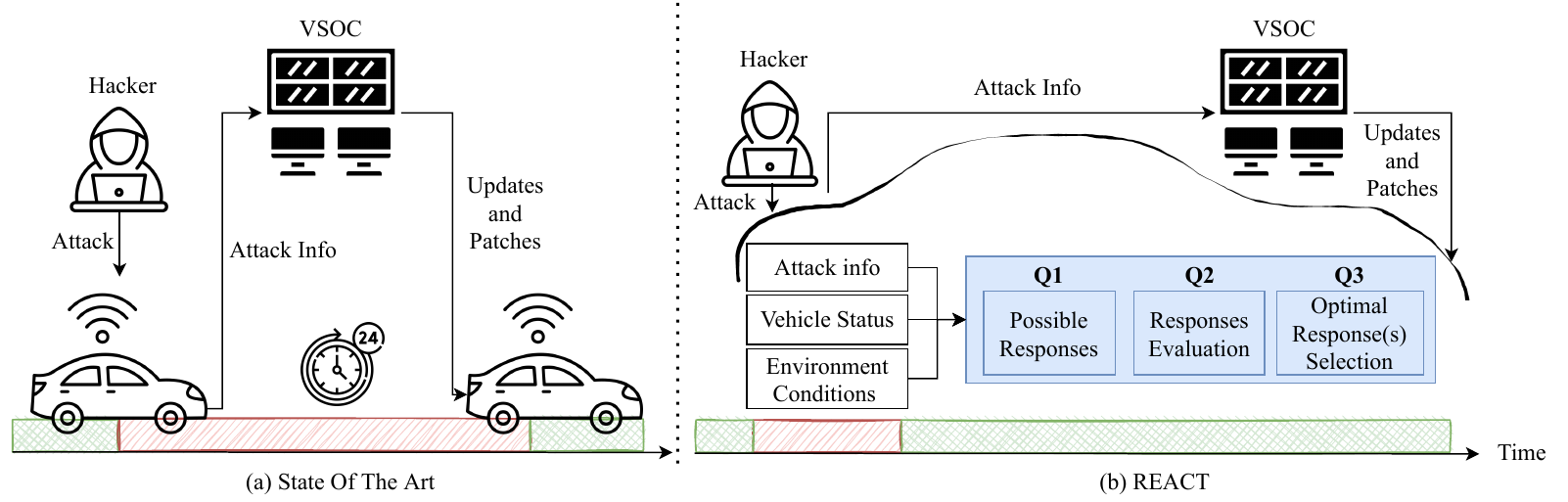}
 	\caption{On the left side, the current vehicle system shares attack information with the VSOC but often has to wait for extended periods to receive necessary security patches and updates. This waiting period puts the vehicle in a malicious status (red, diagonal lines). On the right side, the vehicle can select and implement security solutions to avoid the long waiting time for security patches and updates and return to normal status (green, cross diagonal lines).}
 	\label{fig:contributionfigure}
 \end{figure*}
 
Recently, there has been a growing interest in addressing the security threats that may target smart vehicles. For instance, the ISO 21434 \cite{international2021iso} standard has been introduced, with a significant portion dedicated to the development of threat analysis and risk assessment methodologies. Moreover, the field of intrusion detection and prevention in the automotive domain has witnessed extensive research, leading to various avenues for research \cite{kim2021cybersecurity}. However, despite these efforts, the number of attacks targeting smart vehicles continues to rise \cite{upstream2022}. This is to be expected, as security is not absolute, and we must acknowledge that complete prevention of all security threats may not be attainable. 
Therefore, greater emphasis should be placed on defining \textit{how the system should behave when confronted with such unavoidable attacks.}

The cybersecurity incident response is an integral aspect of security management, as outlined in ISO/SAE 21434 within the operational and maintenance clause \cite{international2021iso}. 
Based on the standard, this process aims to provide remedial actions and updates, which may involve post-development changes to address security vulnerabilities. The process necessitates the vehicle to share cybersecurity information about the vulnerability that triggered the cybersecurity incident response.
Being part of the ISO/SAE 21434,  it is now imperative that manufacturers comply with new regulations by having a cybersecurity management system that oversees the cybersecurity activities and processes in the product life-cycle. 
To achieve this, \acp{VSOC} will be utilized to support monitoring \cite{barletta2023v, sembera2020iso, olt2019establishing}. 
Such \acp{VSOC} will employ expert teams that continuously analyze data collected from all connected vehicles, enabling automakers to swiftly and efficiently address security incidents \cite{olt2019establishing}.  
Although it's arguable that numerous tasks within a \ac{VSOC} could be automated, the challenge of scalability persists, especially considering the extensive fleet of connected vehicles and the immense data volumes accumulated by each vehicle, reaching terabytes \cite{wright2021autonomous}. 
The transfer and processing of such data turn out to be significant issues, particularly in urban areas with hundreds of cars per vicinity, leading to bottlenecks. Additionally, the connectivity itself could be an attractive target for attackers. 
In this context, the integration of \acp{VSOC} into the smart vehicle ecosystem demands solutions for addressing connectivity challenges between vehicles and the \ac{VSOC}, as well as managing privacy concerns tied to shared data \cite{hamad2023security}. 

Finally, and more importantly, there is a need to ensure a near-real-time response to security attacks. Taking into account the need for a human in the loop, as well as the latency introduced by high-volume shared data and communication between the vehicles and the \ac{VSOC}, achieving a near-real-time response seems unrealistic. 
This perspective is supported by the European Union Agency for Cybersecurity (ENISA), which has cautioned that responding to high-criticality attacks could potentially take days or even weeks \cite{european2019enisa}.   
The scenario of extended waiting presents a dilemma, with two options, each having its own disadvantages. Allowing a vehicle to operate with a compromised component due to extended waiting for a security update is far from the ideal situation. Alternatively, suspending the compromised component until the security update is received might not be the best course of action either, particularly if the component plays a crucial role in operations.


\textit{Contributions:}
Therefore, there is a need for vehicles to be equipped with the capability to swiftly respond to cyberattacks. However,  having such a capability requires the answering of three main questions (see \figurename~\ref{fig:contributionfigure}): \textbf{Q1:} What are the possible responses that can be taken? \textbf{Q2:} 
What factors need to be considered when evaluating these responses?
\textbf{Q3:} How to select one or more of these responses at the run-time based on the responses' evaluation?   
This paper aims to address these questions by investigating and categorizing potential responses according to the impact of various cyber attacks to which each response aims to react. Additionally, the paper presents a dynamic risk assessment and cost evaluation for attacks and responses, utilizing given data such as attack information and vehicle status. This assessment supports the selection of suitable responses. Furthermore, the paper explores different approaches for response selection, conducts comparisons, and identifies those best suited for automotive systems. Lastly,  the paper introduces an incident response system,  referred to as REACT, evaluates it using two attack scenarios, and discusses both the quality of the responses it generates and its overall efficiency. 
In summary, the main contributions of this paper are as follows: 

\begin{itemize}
	\item We conduct a comprehensive review of existing intrusion response strategies for IT systems and map them to automotive systems, considering the unique characteristics of automotive attacks and automotive system architectures (see \cref{sec:possibleresponse}).
	\item We propose a novel method for calculating the cost and response benefits by extending existing risk assessment approaches specific to automotive systems (see  \cref{sec:TARA_formalizationofintrusionsandresponses}).
	 \item We explore a range of algorithms for selecting appropriate responses, conduct comparative analyses, and identify the most suitable algorithms for automotive systems, proposing their adoption to enhance automotive security (see \cref{sec:possiblealgorithm}). 
	 
	 \item We introduce REACT, a  comprehensive automotive \ac{IRS}, and provide an open-source prototype\footnote{https://github.com/mohammadhamad/REACT} (see \cref{sec:AutoIRS}).
	 
	\item We demonstrate the feasibility and applicability of the proposed automotive \ac{IRS} through evaluations using embedded platforms and two attack scenarios. Findings indicate that the system can adapt to different scenarios, makes response selections quickly (average 30 ms for the worst-case algorithm), has low memory overhead, and dynamically adjusts system parameters  (see \cref{sec:eval}).
\end{itemize}

%% file: sections/frm.tex
{
	\color{black}

\section{Response Strategies}
\label{sec:possibleresponse}
The purpose of this section is to address the first question (\textbf{Q1}) about possible response strategies. To do so, it is critical to have a deep understanding of the system as well as the potential attacks and threats it may face. Therefore, this section introduces the design of an automotive reference architecture, discusses the potential threats that may arise, and provides a comprehensive summary of the different response strategies that can be utilized to mitigate these attacks. 

\subsection{Automotive Reference Architecture}
In order to understand how \ac{IRS} can be integrated into modern vehicles and the potential responses they can provide, it is essential to first understand their system architecture.  
\figurename~\ref{fig:bordnet_architectureIEEE} presents a  generic, realistic and comprehensive reference architecture that can be found in modern vehicles. 
It is notable that a modern vehicle includes \textit{highly interconnected} subsystems. 
The figure also shows how modern vehicles have many  \textit {embedded devices}, known as \acp{ECU},   which are \textit{distributed} allover the vehicle, communicating among themselves via different types of networks such as CAN, Flexray and Ethernet.  
These \acp{ECU} are grouped in different domains or zones based on the functionality  such as  infotainment, \ac{ADAS},  powertrains, etc.  
Besides \acp{ECU}, modern vehicles are equipped with many sensors (e.g., cameras, LiDAR, etc.), advanced communication technology for connecting with the external world, and diagnostic ports (e.g., OBD-II) that collectively form a significant attack surface for different types of attacks and threats \cite{checkoway2011comprehensive}. 
The unrestricted or/and uncontrolled interaction among all those components puts the whole system in danger. Attackers could launch a \textit{stepping-stone} attack  \cite{48}, where they compromise a non-critical \ac{ECU} with weaker security (e.g., the infotainment system), in order to gain control of a more crucial one (e.g., engine control). 
All these characteristics of the vehicle architecture suggest that any proposed \ac{IRS} should take into account the constrained resources and the highly interconnected and distributed nature of a vehicular system.

\begin{figure}[t]
	\includegraphics[width=0.9\columnwidth]{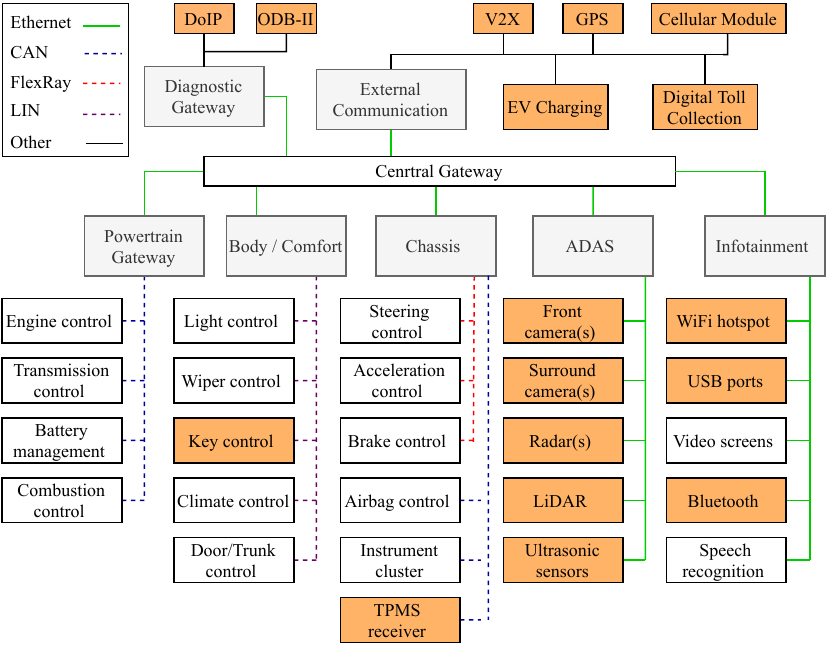}
	\caption{Reference vehicle architecture with possible attack surfaces (orange).}
	\label{fig:bordnet_architectureIEEE}
\end{figure}

\subsection{Threats and Attacks}
\label{sec:TARA_intrusionresults}
\ac{TARA}, an essential component of ISO 21434, is employed as a systematic way to identify and assess cybersecurity threats and risks in the automotive industry, facilitating the implementation of effective mitigation strategies. Since \ac{TARA} does not dictate a specific method to identify threats, various methods have been proposed, such as STRIDE \cite{51}, SAVTA \cite{62}, attack trees \cite{henniger2009securing, 61}, and many others \cite{66}. Following the methodology of \ac{TARA}, these methods provide a comprehensive list of threats and attacks that may target the vehicular system and offer preventive measures. However, they do not address the reactive measures required for an automotive \ac{IRS}.

Using the list of threats and attacks to create a response for each of them seems to be not ideal due to several challenges, including the large number of attacks and the requirements for precise information about each attack, which must be provided by the \ac{IDS}. This challenge becomes evident when considering Zero-Day attacks, where information about such attacks may not be available to the \ac{IRS} at the time of detection by the \ac{IDS}. Even if an anomaly-based \ac{IDS} shares some information about the attack pattern with the \ac{IRS}, a response solely based on known attack patterns may not sufficiently react to these Zero-Day attacks.
Therefore, the most effective approach is to enable the \ac{IRS} to understand the situation it aims to respond to. This involves focusing on the impact or outcome of different attacks rather than solely on the attacks themselves.


To achieve that, we have developed a model, illustrated in  \figurename~\ref{fig:intrusion_results}, which represents the actual results of intrusions collected from various research works. The model encompasses five main attack outcomes, each of which can result from multiple types of attacks. Examples of these attacks are depicted in the outer nodes of \figurename~\ref{fig:intrusion_results}. 
Also, to reflect the outcome of stepping-stone attacks, the model links the different outcomes to demonstrate that certain attacks may cause a series of results.
The five attack outcomes are:

	\begin{figure}[t!]
	\centering
	\includegraphics[width=0.75\columnwidth]{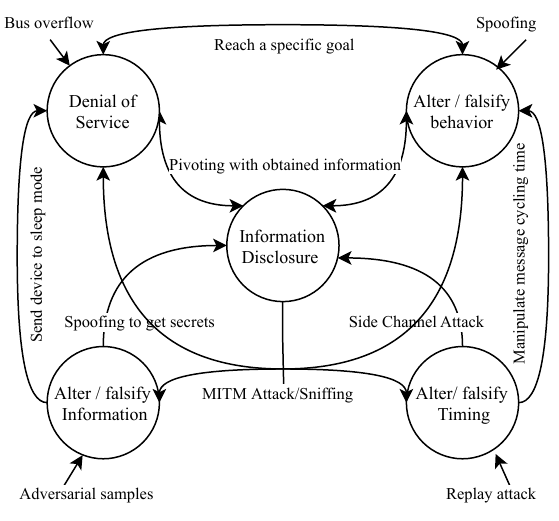}
	\caption{Classification of intrusion results and examples of attacks for each possible intrusion
	result.}
	\label{fig:intrusion_results}
\end{figure}

\begin{itemize}
	\item \textit{Falsify / Alter Information:} 	
	Different attacks have the potential to modify information on a bus or within an \ac{ECU}. It is important to note that not every alteration of information automatically results in undesirable behavior. For instance, adversarial samples \cite{81}, such as incorrect classifications of objects detected by a camera, may not necessarily lead to incorrect behaviors.
	
	\item \textit{Falsify / Alter Timing:} 
This outcome typically occurs as a result of attacks targeting the communication buses of the vehicle \cite{22, 15} or the real-time tasks on the \acp{ECU} \cite{hamad2018prediction}. 
	\item \textit{Information Disclosure:}  
This outcome is the result of attacks, such as spoofing, eavesdropping, and others, that aim to allow attackers to gain unauthorized access to sensitive information exchanged during communication or stored within the \acp{ECU} \cite{cui2019review}.

	\item \textit{System Unavailability:}  This outcome typically occurs as a result of Denial of Service (DoS) attacks that aim to cause a loss of availability for a specific component or subsystem in the vehicle \cite{palanca2017stealth}. Such attacks can lead to severe damage to the system, especially if they target high-critical components \cite{alrefaei2022survey}.
	\item \textit{Falsify / Alter behavior:} This outcome is the result of tampering attacks that specifically target the components, data, or parameters of a system with the intention of altering the system's intended behavior and achieving unauthorized or malicious outcomes \cite{miller2015remote}. While this intrusion outcome may appear similar to falsify/alter information, the key distinction is that in falsify/alter information attacks, the goal is to tamper with the information itself without the explicit method of changing the system's behavior, even though it may indirectly lead to such changes.
\end{itemize}

\subsection{Response Possibilities}\label{sec:TARA_responsepossibilities}

\begin{table*}[t!]
	\caption{Classification of generic responses to intrusion results.}
	\label{tab:responses}
	\begin{tabular*}{0.99\textwidth}{@{\extracolsep{\fill}} p{0.2\textwidth}p{0.77\textwidth}@{}}
		\toprule
		\textbf{Intrusion Result}      & \textbf{Response Index.  Response}  \\
		\midrule 
		Falsify / Alter Timing      & 
		 \textbf{1}. Use of redundant information  \cite{1}, 
		 \textbf{2}. Correction of timing ~\cite{3, 88}, \textbf{3}. Force additional authentication ~\cite{13}, 
		 \textbf{4}. Restart the device/system ~\cite{6}, \textbf{5}. Change settings ~\cite{11}, 
		 \textbf{6}. Redirect traffic ~\cite{11}, 
		 \textbf{7}. Re-initialization ~\cite{43}\\ 
		Falsify / Alter Information & 
		\textbf{1}. Use of redundant information (Reallocation) \cite{1}, 
		\textbf{3}. Force additional authentication ~\cite{13}, 
		 \textbf{4}. Restart the device/system ~\cite{6}, \textbf{8}. Create a backup \cite{49}, 
		 \textbf{5}. Change settings ~\cite{11}, 
		 \textbf{7}. Re-initialization ~\cite{43},  
		 \textbf{9}. Correct protocol specification faults ~\cite{24}, 
		 \textbf{10}. Split or merge functions \cite{32}\\ 
		Information Disclosure      & 
		\textbf{11}. Issue authentication challenges \cite{3}, 
		\textbf{12}. Re-enforce access control \cite{4}, \textbf{3}. Force additional authentication \cite{13}, 
		\textbf{13}. Introduce a honeypot \cite{4}, 
		\textbf{4}. Restart the device/system ~\cite{6}, \textbf{14}. Modify firewall \cite{11}, 
		\textbf{6}. Redirect traffic \cite{11}, 
		\textbf{10}. Split or merge functions \cite{32}, \textbf{7}. Re-initialization ~\cite{43}, 
		\textbf{15}. Network isolation \cite{88}  \\ 
		System Unavailability          & 
		\textbf{1}. Use of redundant information (Reallocation) \cite{1}, 
		\textbf{12}. Re-enforce access control \cite{4}, \textbf{13}. Introduce a honeypot \cite{4}, 
		\textbf{4}. Restart the device/system (source or destination) ~\cite{6}, 
		\textbf{14}. Modify firewall \cite{11}, 
		\textbf{6}. Redirect traffic \cite{11}, 
		\textbf{10}. Split or merge functions \cite{32}, 
		\textbf{7}. Re-initialization ~\cite{43}, 
		\textbf{16}. Limit resources of the attacker \cite{49}, 
		\textbf{17}. Safe mode \cite{2}\\ 
		Falsify / Alter Behavior   & 
		\textbf{1}. Use of redundant information (Reallocation) \cite{1}, 
		\textbf{18}. Correction of behavior \cite{3}, 
		\textbf{9}. Correct protocol specification faults ~\cite{24}, 
		\textbf{3}. Force additional authentication \cite{13}, 
		\textbf{19}. Restart the miss-behaving system \cite{6}, 
		\textbf{5}. Change settings ~\cite{11}, 
		\textbf{10}. Split or merge functions \cite{32}, 
		\textbf{7}.  Re-initialization of the miss-behaving device ~\cite{43}, 
		\textbf{17}. Safe mode \cite{2}, 
		\textbf{8}. Create a backup \cite{49}\\ 
		General   & 
		\textbf{20}. Isolation \cite{1}, 
		\textbf{21}. Limit communication of malicious system \cite{1}, 
		\textbf{22}. Drop packets \cite{6}, 
		\textbf{23}. Trace communication \cite{1}, 
		\textbf{24}. Introduce additional logging \cite{13}, \textbf{25}. Block network traffic \cite{4}, 
		\textbf{26}. Kill process \cite{1}, 
		\textbf{27}. Reduce trust level of the source \cite{1}, 
		\textbf{28}. Perform a security auditing \cite{2}, \textbf{29}. Request / Perform software update \cite{3}, 
		\textbf{30}. Notify \ac{SOC} / administrator~\cite{10, 4}, 
		\textbf{31}. No action \cite{10}, 
		\textbf{32}. Adapt parameters for \ac{IDS} \cite{45}, \textbf{33}. Warn / inform other \acp{ECU} \cite{26, 1} \\ 
		\bottomrule
	\end{tabular*}
\end{table*}

After classifying the outcome of the attack, it becomes easier to determine which responses can be used to address that particular outcome and handle the attacks that cause it. In order to do so, we have examined typical responses discussed in both the automotive and non-automotive domains. It should be noted that while some research papers in the automotive domain have discussed the need for responses to certain attacks, there is currently no comprehensive research that lists and classifies all possible  responses. Furthermore, it is important to consider that some of the responses we collected were originally designed for computer networks and may not be directly applicable to automotive bus systems due to the lack of specific security mechanisms~\cite{88}. For example, response actions such as IP address changes or port blocking~\cite{13} are highly specific to Ethernet and higher protocols such as IP, and therefore have limited suitability for certain aspects of communication in vehicles. To address this challenge, we have defined a list of generic responses that are specific enough to be applied in an automotive IRS, while also being adaptable to constrained and potentially insecure devices. Table~\ref{tab:responses} provides an overview of the different responses based on the identified attack outcomes. In addition, we have included a \quotes{General} category that encompasses responses applicable to all five categories. For more detailed information about each response, please refer to the respective sources cited in Table~\ref{tab:responses}.

	}

%% file: sections/responseevalaution.tex
\section{Dynamic Cost and Impact Evaluation}
\label{sec:TARA_formalizationofintrusionsandresponses}

In this section, we will address \textbf{Q2} by outlining the key factors required to enable the selection of the most effective response by the \ac{IRS}. These factors can be categorized into two groups: \textit{intrusion-related factors}, which pertain to the attack's impact and risk, and \textit{response-related factors}, which concern the cost and benefit of the chosen response.

	
\subsection{Intrusion-Related Factors}
	\label{sec:TARA_formaldescriptionofintrusions}
\subsubsection{Intrusion Properties}
For each detected intrusion, the following properties need to be determined:   
\begin{itemize}
	\item \textit{Source of the intrusion:} This represents the component from which the attack was launched. Referring to the automotive reference architecture depicted in \figurename~\ref{fig:bordnet_architectureIEEE}, sources can include entities from the attack surface as well as external attackers targeting any of these components.
	\item \textit{Destination of the intrusion:} The attacked entity can be described as the destination of the intrusion. This could be \acp{ECU}, sensors, or bus systems. 
	\item \textit{Intrusion result:} This refers to one of the outcomes that were previously defined in Subsection~\ref{sec:TARA_intrusionresults}. Similar to the source and destination of an intrusion, this information is also provided by an \ac{IDS}. 
	\item \textit{Intrusion impact}: This information serves to depict the impact of the intrusion on the system and is essential for evaluating the risks during the attack.
\end{itemize}

\subsubsection{Dynamic Attack Impact Assessment}
To assess the potential risks associated with an intrusion, it is necessary to understand the impact of the attack and the likelihood of its occurrence \cite{international2021iso, lautenbach2021proposing}. 
To calculate the impact of the intrusion,  many methods were already adopted such as HEAVENS \cite{islam2016risk}. HEAVENS classifies the impact of a given threat based on four metrics~ \cite{63, 66}:
	
	\begin{enumerate}
		\item Safety impact, denoted as $S$ with $S \in \{0,10,100,1000\}$
		\item Financial impact, denoted as $F$ with $F \in \{0,10,100,1000\}$
		\item Operational impact, denoted as $O$ with $O \in \{0,1,10,100\}$
		\item Privacy impact, denoted as $P$ with $P \in \{0,1,10,100\}$
	\end{enumerate}
	
	In the original HEAVENS method, the overall impact $I$ is calculated as a sum of the four single impacts as depicted in Equation~\ref{eq:HEAVENS}~\cite{63}.
	
	\begin{equation}
	\label{eq:HEAVENS}
	I=S+F+O+P
	\end{equation}

One issue with the impact calculation, as presented in Equation~\ref{eq:HEAVENS}, is the overemphasis on safety and financial parameters. This skewed emphasis not only complicates the comparison and independent evaluation of the four metrics but also renders it unsuitable for an automotive \ac{IRS}. 
In the automotive context, safety and operational considerations typically outweigh financial and privacy-related aspects for most automotive functions.
Considering the aforementioned issue, we propose normalizing all possible values to ${0, 1, 10, 100}$, representing no, low, medium, or high impact for each of the four metrics in HEAVENS.

Another limitation of the current risk assessment methods, including HEAVENS, is their failure to account for dynamic environmental factors, such as run-time context, operational status, and the surrounding environment. This gap may arise because HEAVENS is primarily applied during the design phase, making it somewhat oblivious to run-time conditions. 
To address this challenge and enhance the method's applicability for use within automotive \ac{IRS}, we introduce a new metric termed "Environment," denoted as $E$. This metric, $E$, encompasses dynamic factors that are crucial for assessing intrusion impact \cite{1}. Potential inputs that can be used to derive the environmental parameter  $E$ include vehicle speed, road conditions, the proximity of nearby objects, and more. These parameters can exert significant influence, as a single intrusion may yield different impacts depending on physical and environmental considerations.

The final enhancement option for the HEAVENS method involves the capability to dynamically adjust the assessment of intrusion impact. Following a successful intrusion response, it may become evident that the stored parameters for $S$, $F$, $O$, $P$, and $E$ require a different representation. 
HEAVENS currently confines impact values to ${0, 1, 10, 100}$, and a simple adjustment to a new value could result in significant over-representation. 
To address this issue, introducing weights for each of the five evaluation metrics ($w_S$,  $w_F$, $w_O$, $w_P$, and $w_E$) offers a valuable mechanism for accommodating learning and adaptation processes. The optimization proposals discussed earlier to transform the calculation of intrusion impact using the HEAVENS method into a dynamic process lead to  Equation~\ref{eq:HEAVENS_adapted}.

\begin{equation}
	\label{eq:HEAVENS_adapted}
	I=w_S\cdot S+w_F\cdot F+w_O\cdot O+w_P\cdot P+w_E\cdot E
		\end{equation}		
		
Utilizing dynamically adjusted static values for $S$, $F$, $O$, and $P$, each incorporating their respective weights, in addition to dynamically acquired values for $E$ along with an adapted static weight. In cases involving specific automotive architectures, the equation can also be applied in a more granular fashion for particular assets. Initial values for all these parameters can be established by security experts, drawing upon their experiential knowledge.
	
The source and destination of the attack are employed to determine the attack's location, aiding in the calculation of the subsequent attack likelihood, especially when considering step-stone attacks, across various parts of the system. This assessment of attack likelihood, in conjunction with the evaluation of attack impact, contributes to the overall risk assessment.
	
\subsection{Response-Related Factors} \label{sec:TARA_formaldescriptionofresponses}
		
	\subsubsection{Response Properties}
	Similar to the intrusion, each response will have five properties that need to be identified:
		\begin{itemize}
		\item \textit{Actual action:} They refer to the actual actions taken in the event of an intrusion. These actions can be selected from those presented in Table~\ref{tab:responses}. 
		\item \textit{Precondition:} Some responses may require preconditions that must be met. These preconditions can be expressed as Boolean expressions and serve as prerequisites to trigger the response. 
		\item \textit{Place of application:} Refers to the location where the response will be implemented. A response can be applied either at the source entity of an intrusion, the destination, or at both locations. 
		\item \textit{Stop condition:} Refers to the condition for which the implemented response should cease. This condition can be related to a specific time~\cite{55}, the successful reestablishment of security policies~\cite{1}, or the necessity for persistent measures~\cite{48}. 
		\item \textit{Cost and benefit of the response:} Refers to the costs and benefits incurred when implementing a response to an intrusion or security incident.  
	\end{itemize}

	\subsubsection{Dynamic response cost and benefit assessment}
When considering the cost of responses, various methods were employed to determine their value in IT systems ~\cite{9}. These methods primarily rely on one of three models: a static cost model that assigns a fixed cost value for each response, a static evaluated cost model that calculates cost using a static function with some adjustment possibilities, or dynamic evaluated cost models that offer fully dynamic evaluation based on real-time data. Each model varies in terms of simplicity, adaptability, and accuracy, catering to different system requirements and scenarios. 

	Statically evaluated cost models provide a valid trade-off between achievable implementation efforts, especially on constrained devices similar to the ones used in automotive systems, and plausible results. These models maintain a static approach to calculating response costs, even though the actual cost values may vary. Various metrics for calculating response costs are mentioned in current literature. 
	The first metric evaluates the impact of the response on availability~\cite{9}. Availability's impact is represented as $A \in {0, 1, 10, 100}$ to ensure consistency with intrusion metrics. 
	The second metric, describing the response cost, assesses its effect on the performance of the (sub)system~\cite{9}, similar to the deployment cost of countermeasures~\cite{59}. This metric is denoted as $Perf \in {0, 1, 10, 100}$ to maintain a uniform scale with the impact of the response on availability.

	To achieve results similar to the adapted HEAVENS method described in \cref{sec:TARA_formaldescriptionofintrusions}, a comparable equation can be employed to calculate the cost ($c$) of a response. By adopting specific weights ($w_A$ and $w_{Perf}$) for the impact on availability and performance along with their actual values ($A$ and $Perf$), the response cost can be computed as shown in Equation \ref{eq:responsecost}.
	This approach results in a highly adaptable method for calculating the response cost. While the initial values for $A$ and $Perf$ can be manually determined, they can also be adjusted over time. The specific weights offer a means to introduce a learning component within the mathematical framework.

	\begin{equation}
	\label{eq:responsecost}
	c=w_A\cdot A+w_{Perf}\cdot Perf
	\end{equation}

	Likewise, the adapted HEAVENS method introduced in  \cref{sec:TARA_formaldescriptionofintrusions} can be repurposed for evaluating the benefit of a response, with the exception of the environmental parameter $E$ and its associated weight $w_E$. While HEAVENS assesses intrusion impact using four metrics, these same metrics can be employed to quantify the benefits in these four categories when assessing response value. By employing identical value possibilities with $S, F, O, P \in {0, 1, 10, 100}$, a corresponding benefit value can be determined. The calculation of the benefit ($b$) for each response option, as shown in Equation~\ref{eq:HEAVENS_responses}, is derived from Equation~\ref{eq:HEAVENS_adapted}.
	
	\begin{equation}
	\label{eq:HEAVENS_responses}
	b=w_S\cdot S+w_F\cdot F+w_O\cdot O+w_P\cdot P
	\end{equation}

For each response option classified in Table~\ref{tab:responses}, the cost calculated using Equation~\ref{eq:responsecost} and the benefit determined using Equation~\ref{eq:HEAVENS_responses} must be applied, and preconditions must be established. Initial values for $S$, $F$, $O$, $P$, $A$, and $Perf$, along with their respective weights, can be assigned by security experts and subsequently updated either manually or through learning algorithms within an \ac{IRS}. Similar to the impact calculation of intrusions, these weights can be adjusted to improve the accuracy of the model.

%% file: sections/selection.tex
{
	\color{black}

\section{Optimal Selection Algorithms}
\label{sec:possiblealgorithm}
In this section, we will address the third question \textbf{Q3}, by exploring numerous potential methods for selecting response strategies (\cref{sec:posiblealgo}), compare these approaches and provide a rationale for our chosen strategy (\cref{sec:compare}), and describe how to adopt the selected strategies (\cref{sec:responseadption}). 

\subsection{Possible Algorithms}
\label{sec:posiblealgo}
To determine the best method for selecting appropriate responses, we explore various algorithms and solutions used in \textit{non-automotive domains} and compare them to identify the most suitable one that can be implemented within the vehicle system. 
Several surveys, such as \cite{8169023, irssurvey22, BASHENDY2023102984}, provide valuable insights into response selection approaches in non-automotive domains, making them worth investigating for more comprehensive details.

\subsubsection{\ac{SAW}}
SAW \cite{fishburn1967additive} is the simplest and most often used method. 
The basic concept of this method is to find a preference value ($p$) for each possible response, and then select the response with the highest preference value as the best option. 
To illustrate how this method works, let us assume that we have $n$ possible responses ($\mathcal{R} =\{r_1, r_2, \dots, r_n\}$) and $m$ criteria ($\mathcal{CR} =\{cr_1, cr_2, \dots, cr_m\}$) that will be used as a reference for evaluating the responses. 
Each criterion will be assigned a weight $w_j$ where $\sum_{j=1}^m w_j = 1$. To calculate the preference values, a normalized decision matrix is first created, where each element of the matrix is normalized based on the nature of the criterion, whether it is a cost or benefit, as shown in Equation \ref{eq:SAW_normalize}.
	\begin{equation}
		\label{eq:SAW_normalize}
		\alpha_{ij}= 
		\begin{cases}
		\frac{v_{i,j}}{\max_{i}(v_{i,j})},& \text{if criterion $cr_j$ is a benefit}\\
		\frac{\min_{i}(v_{i,j})}{v_{i,j}},& \text{if criterion $cr_j$ is a cost}
		\end{cases}
		\end{equation}
		
where $v_{i,j}$ is the performance value of the response $r_i$ when it is evaluated in terms of criterion $cr_j$. 
The preference value ($p_i$) of response $r_i$ is then obtained by calculating the weighted sum of the normalized performance values using Equation \ref{eq:SAW_weight}.
\begin{equation}
\label{eq:SAW_weight}
p_i = \sum_{j=1}^m w_j \cdot \alpha_{ij}
\end{equation}
Finally, the response $r_i$ with the highest preference value ($p_i$) is considered as the best selection response.


\subsubsection{Linear Programming (LP)} 	
LP is a mathematical technique that can be employed to select optimal responses \cite{39}. 
LP can be used to find the best combination of responses that \textit{maximizes} or \textit{minimizes} a certain objective function. To illustrate the workings of this method, let's consider a scenario where we have $n$ possible responses ($\mathcal{R} = {r_1, r_2, \dots, r_n}$).  
The optimization of the objective function can be as in Equation \ref{eq:lp_objectivefunction}. 

\begin{equation}
\label{eq:lp_objectivefunction}
\sum_{i=1}^{n}x_is_i\rightarrow \max or \min
\end{equation}
where $x_i$ represents a criterion related to the response $r_i$ and $\overrightarrow{s}$ be a vector of binary decision variables, where  $s_i$ is equal to \SI{1}{}, it indicates that the corresponding response $r_i \in \mathcal{R}$ will be executed. Conversely, if $s_i$ is equal to \SI{0}{}, it signifies that the response $r_i \in \mathcal{R}$ will not be executed. 
The optimization problem typically includes \textit{constraints} to ensure the selection process adheres to specific conditions or limitations.


\subsubsection{Game-Theoretic Algorithm}
Another mathematical method to determine optimal responses against cyber attacks is game-theoretic algorithms \cite{32, 116, 117}.  
In the game-theoretic approach, the attacker and the \ac{IRS} are modeled as two players. Each player has a set of actions available to them, such as different attack strategies $\mathcal{A}=\{a_1, a_2, \dots, a_k\}$ for the attacker and response strategies $\mathcal{R}=\{r_1, r_2, \dots, r_n\}$ for the IRS. 
The goal of the \ac{IRS} is to select the optimal response to the attack at a given time. One way to achieve that is by minimizing the maximum damage of the attack: $\min_{r_i \in\mathcal{R}}(\max_{a_i \in\mathcal{A}}(U(r_i,a_i)))$ where $U(r_i,a_i)$ represents the utility function for the \ac{IRS} when the attacker chooses attack $a_i$ and the \ac{IRS}  responds with response $r_i$. 


\subsubsection{AI-based mechanisms} 
Many AI-based mechanisms were used to support the dynamic selection of the response such as Genetic Algorithms \cite{40}, Convolutional Neural Networks \cite{47}, Supervised machine learning \cite{54},  Q-Learning \cite{8685487}, and many more \cite{rose2022ideres}. Using any of these AI models usually requires many steps including data collection and pre-processing, feature extracting, model training, and feedback loop to improve the quality of the selected responses.

\subsubsection{Other Methods} There are alternative mathematical approaches to \acp{IRS} that are not derived from general mathematical problems. One example is REASSESS~\cite{111} that uses human-evaluated metrics and prior responses to select optimal responses. While it offers simplicity, this reliance on human evaluation can lead to inaccurate assumptions. Its mandatory learning behavior is unsuitable for automotive systems, and it lacks the option for flexible learning to enhance responses, requiring a well-established feedback loop. 
Another simpler approach is the cost-sensitive generic framework~\cite{112, 114}, which includes steps like defining operational costs, ranking responses using a weighted sum method, and selecting the best response with an intrusion matrix. However, its reliance on static value assignments and sensitive parameters, typically defined by human experts, can make objective assessment challenging and results in potentially harmful responses.

\begin{table*}[t!]
	\caption{Comparison of the different response selection methods}
	\label{tab:algorithm_compare}
	\begin{tabular*}{0.99\linewidth}{@{\extracolsep{\fill}} p{0.2\linewidth}p{0.33\linewidth}p{0.43\linewidth}@{}}
		\toprule
		\textbf{Method}  & \textbf{Benefits} & \textbf{Drawbacks} \\
		\midrule
		\textbf{\ac{SAW}}   & 
		\begin{tabular}[t]{@{}p{\linewidth}@{}}
			\good{+} Simplicity and lightweight operators  \\
			\good{+} Suitable for constrained devices  \\
			\good{+} Polynomial run-time 
		\end{tabular}
		& \begin{tabular}[t]{@{}p{\linewidth}@{}}
			\bad{-} Adapted methods for accuracy increase complexity \\ 
			\bad{-} Reliance on subjective parameters
		\end{tabular}
		\\
		\midrule
		\textbf{LP} 
		& 
		\begin{tabular}[t]{@{}p{\linewidth}@{}}
			\good{+} Flexible structures\\
			\good{+} Typically polynomial run-time\\
			\good{+} Existing libraries for solvers
		\end{tabular}
		& \begin{tabular}[t]{@{}p{\linewidth}@{}}
			\bad{-} Higher complexity for modeling and calculation\\ 
			\bad{-} Theoretically exponential run-time
		\end{tabular}
		\\  
		\midrule
		\textbf{Game-Theoretic Algorithms}  & 
		\begin{tabular}[t]{@{}p{\linewidth}@{}}
			\good{+} System state consideration\\
			\good{+} Accurate system  representation 
		\end{tabular}
		& \begin{tabular}[t]{@{}p{\linewidth}@{}}
			\bad{-} Very complex models\\
			\bad{-} Computational complexity\\
			\bad{-} Reliance on subjective parameters
		\end{tabular}
		\\
		\midrule
		\textbf{AI-based Solutions}   & 
		\begin{tabular}[t]{@{}p{\linewidth}@{}}
			\good{+} Handle large amount of data \\
			\good{+} Fast response selection\\
		
		\end{tabular}
		& \begin{tabular}[t]{@{}p{\linewidth}@{}}
				\bad{-} Uncertainty of the selected responses\\
				\bad{-} High resource requirements
		\end{tabular}
		\\
		\midrule
		\textbf{Other Methods}    & 
		\begin{tabular}[t]{@{}p{\linewidth}@{}}
			\good{+} Simple mathematical models\\
			\good{+} Typically fast \\
			\good{+} Combination with other methods possible\\
			\good{+} Learning is possible
		\end{tabular}
		& \begin{tabular}[t]{@{}p{\linewidth}@{}}
			\bad{-} Complexity raises with large systems\\
			\bad{-} Human influence has always subjective opinions
		\end{tabular}
		\\
		\bottomrule
	\end{tabular*}
\end{table*}

\subsection{Comparison}
\label{sec:compare}
 Table~\ref{tab:algorithm_compare} summarizes all the advantages and the drawbacks of the five classes of response selection algorithms.  
 
 The primary advantage of \ac{SAW} is its relative simplicity and utilization of lightweight mathematical operators, making it suitable for running on constrained devices with a polynomial run-time, without requiring complex external libraries ~\cite{5}.
 However, the main drawback of \ac{SAW} is the need for an adapted \ac{SAW} method to achieve more accurate results. This often leads to increased complexity and longer run-time compared to the original \ac{SAW}. Another drawback is the dependency on subjective parameters such as specific weights. This dependency can result in highly variable outcomes that may not accurately reflect the system state~\cite{107}.
 
 A major benefit of LP is its ability to formulate a single objective function and multiple constraints, providing an accurate representation of multi-objective optimization problems.
 However, compared to SAW, LP requires complex implementation, resulting in increased computational complexity for large systems~\cite{39}. 
 The run-time of the algorithm depends on the solving method employed, such as the commonly used Simplex algorithm. 
 While the Simplex algorithm has polynomial run-time for \quotes{typical} problems~\cite{109}, it exhibits exponential worst-case run-time in theory~\cite{110}. 
 
 The advantage of game-theoretic approaches lies in their consideration of the system state, resulting in a highly accurate representation of the system. 
 Furthermore, game-theoretic approaches can be deployed in a distributed manner, as highlighted in \cite{116}. 
 A major drawback of this method is the use of highly complex models, which are necessary to determine optimal moves in game-theoretic algorithms. Solving such complex models often requires significant resources and leads to large communication overhead~\cite{116}, making this approach unsuitable for constrained devices.
 Additionally, most models in practice make assumptions or simplifications due to the near-infinite number of possible system states~\cite{32, 116, 117}, as complete modeling of all states is infeasible. 
 
 Using AI-based methods is still limited because of many issues such as the high memory and computation requirements of some of these methods \cite{8791960} and the unrealistic responses that some models can produce (e.g., Genetic Algorithms). Additionally, uncertainty surrounding the outputs of these models limits their adoption.  
 Finally, most of these methods rely on the availability of datasets for model training. However, autonomous vehicles often operate in dynamic and unpredictable environments. When the operating environment significantly deviates from what the AI has learned, it may encounter challenges in adapting effectively or making appropriate decisions.

Finally, while the cost-sensitive generic framework and REASSESS are simple and demonstrate promising in computer and network technologies, adapting them to a highly heterogeneous multi-bus architecture, like the vehicular reference architecture, presents significant challenges.
 

After careful consideration of the factors discussed above, we have chosen to explore the adapted \ac{SAW} method, as well as LP with a focus on both benefit maximization and cost minimization for the design of an automotive \ac{IRS}. The decision to focus on these two methods is based on their relative simplicity, computational efficiency, and their ability to accurately represent multi-objective optimization problems. The remaining algorithm families were assessed but are not pursued further due to reasons such as increased complexity, resource requirements, and limitations in modeling all possible system states.

\subsection{Adopting of SAW and LP}
\label{sec:responseadption}

\subsubsection{Adopting of \acs{SAW}}\label{sec:optimalselectionalgorithm_adaptionofSAW}
To adopt the \ac{SAW} method for automotive \acp{IRS}, we first need to define the criteria $\mathcal{CR}$ that will be used to evaluate each response. For this purpose, we can utilize the HEAVENS parameters, including the cost of a response $c$ (see Equations~\ref{eq:responsecost}) and the benefit of a response $b$ (see Equation~\ref{eq:HEAVENS_responses}). However, using these two parameters still presents some issues that need to be addressed in order to effectively use and adapt \ac{SAW} for valid results.
The first problem arises when using these parameters during the creation of the elements of the normalized decision matrix, as depicted in Equation~\ref{eq:SAW_normalize}. This problem originates from the fact that our modified HEAVENS method allows values of $v_{i,j}$ to be in the set ${0,1,10,100}$ for both criteria (i.e., $c$ and $b$). If $\max_{i}(v_{i,j})=0$ applies, Equation~\ref{eq:SAW_normalize} results in an illegal operation if the criterion is a benefit. Similarly, if the criterion is a cost and $v_{a,j}=0$, Equation~\ref{eq:SAW_normalize} also results in an illegal operation. This issue can be circumvented by using a small value greater than $0$ instead of $0$.
The second problem does not stem from a mathematical perspective but rather from the application of this method in a fully automated \ac{IRS}. Since the \ac{SAW} method only considers criteria $\mathcal{CR}$ from the applicable response set $\mathcal{R}$, it does not take into account the impact $I$ of an intrusion. As a result of this limitation, it is possible that a response incurring high costs may be chosen even for a minor intrusion.  Although this is a significant challenge for the application of \ac{SAW} in \acp{IRS}, this drawback has not been addressed in existing research.

To tackle this problem, it is mandatory to set the preference value $p$ (see Equation \ref{eq:SAW_weight}) into relation with the intrusion impact $I$. 
For each asset $A$ of the vehicle reference architecture and each intrusion result $\mathcal{R}$, a normalized intrusion impact can be calculated. 
Such a normalized intrusion impact must be calculated for each metric $S$, $F$, $O$, $P$ and $E$ of the adapted HEAVENS method in Equation~\ref{eq:HEAVENS_adapted}. This behavior is formulated in Equation~\ref{eq:SAW_normalize_response}.

\begin{equation}
\label{eq:SAW_normalize_response}
\begin{matrix*}[l]
\alpha_{\{S,F,O,P,E\},A,\mathcal{R}}=  \\ \\
\begin{cases}
\frac{w_{\{S,F,O,P,E\},A,\mathcal{R}}\;\cdot\; v_{\{S,F,O,P,E\},A,\mathcal{R}}}{\sum_{|\mathcal{R}|}^{}(w_{\{S,F,O,P,E\},A}\;\cdot\; v_{\{S,F,O,P,E\},A})},& \text{if } \sum_{|\mathcal{R}|}^{}(w_{\{S,F,O,P,E\},A}\;\cdot\; v_{\{S,F,O,P,E\},A})\neq 0\\
0,& \text{otherwise}
\end{cases}
\end{matrix*}
\end{equation}

Similar to Equation~\ref{eq:SAW_weight}, a weighted sum must be calculated. But, since the individual weights $w$ are already included in Equation~\ref{eq:SAW_normalize_response}, a simple summation over all metrics $S, F, O, P$ and $E$ of the adapted HEAVENS method is sufficient. This sum will be set into relation with the preference value of the responses from Equation~\ref{eq:SAW_weight}, such that the response $r_i$ with the highest preference value $p$ will be used, which is below the sum of all normalized HEAVENS values as depicted in Equation~\ref{eq:SAW_getResponse}.

\begin{equation}
\label{eq:SAW_getResponse}
\text{best response}=\max\left \{ p_{i}\;|\;p_{i}<\rho\cdot \sum_{l\in\{S,F,O,P,E\}}^{}\alpha_{l,A,\mathcal{R}}\right \}
\end{equation}

The parameter $\rho$ in Equation~\ref{eq:SAW_getResponse} is a parameter to adjust larger deviations in the order of magnitude between the sum of the normalized HEAVENS and the preference value $p$. 

\subsubsection{Adopting of Linear Programming}\label{sec:optimalselection_adaptionoflinearprogramming}
The first step to adopt the LP is defining the objective function. For the set of possible responses $\mathcal{R}$,  it is possible to define two different objective functions:
\begin{itemize}
	\item The first option of an objective function follows the principle of maximum benefit as depicted in Equation~\ref{eq:adoptLP_maxbenefit}. The goal is to solve the binary decision vector $\overrightarrow{s}$ to maximize the benefit $b$. 
	Although this can lead to very good solutions, it is possible that the best executable response is not found immediately since preconditions of identified responses are not satisfied. 
	\begin{equation}
	\label{eq:adoptLP_maxbenefit}
	\sum_{i=1}^{|\mathcal{R}|}s_ib_i\rightarrow \max
	\end{equation}
	\item The second option of an objective function follows the minimum cost principle and is comparable to existing \acp{IRS}~\cite{39, 43}.  Equation~\ref{eq:adoptLP_mincost} therefore leads to more conservative responses since the cost $c$ will be minimized and the benefit $b$ of a response is not considered. A drawback is that the identified solution inside $\overrightarrow{s}$ might not heal the system completely and another try might be necessary.
	\begin{equation}
	\label{eq:adoptLP_mincost}
	\sum_{i=1}^{|\mathcal{R}|}s_ic_i\rightarrow \min
	\end{equation}
\end{itemize}

For both objective functions from Equation~\ref{eq:adoptLP_maxbenefit} and~\ref{eq:adoptLP_mincost} the same constraints must be satisfied for a response to qualify for execution. Existing constraints of \acp{IRS} using LP ~\cite{39, 43} are not suitable for an automotive \ac{IRS}. Because of that, specific constraints must be elaborated:

\begin{enumerate}
	\item The cost $c$ of the response must be below the impact $I$ of the detected intrusion~\cite{39}.  Equation~\ref{eq:adoptLP_constraint1} depicts this first constraint.
	\begin{equation}
	\label{eq:adoptLP_constraint1}
	\sum_{i=1}^{|\mathcal{R}|}s_ic_i< I
	\end{equation}
	\item Only one response can and must be executed as depicted in Equation~\ref{eq:adoptLP_constraint2}.
	\begin{equation}
	\label{eq:adoptLP_constraint2}
	\sum_{i=1}^{|\mathcal{R}|}s_i=1
	\end{equation}
\end{enumerate}

It is additionally necessary that $\overrightarrow{s}$ is a binary vector, leading to the variable definition $s_i\in \{0,1\}$.

%% file: sections/sysarch.tex
\section{Proposed Automotive IRS}
\label{sec:AutoIRS}

\begin{figure*}[t!]
	\centering	\includegraphics[width=0.99\textwidth] {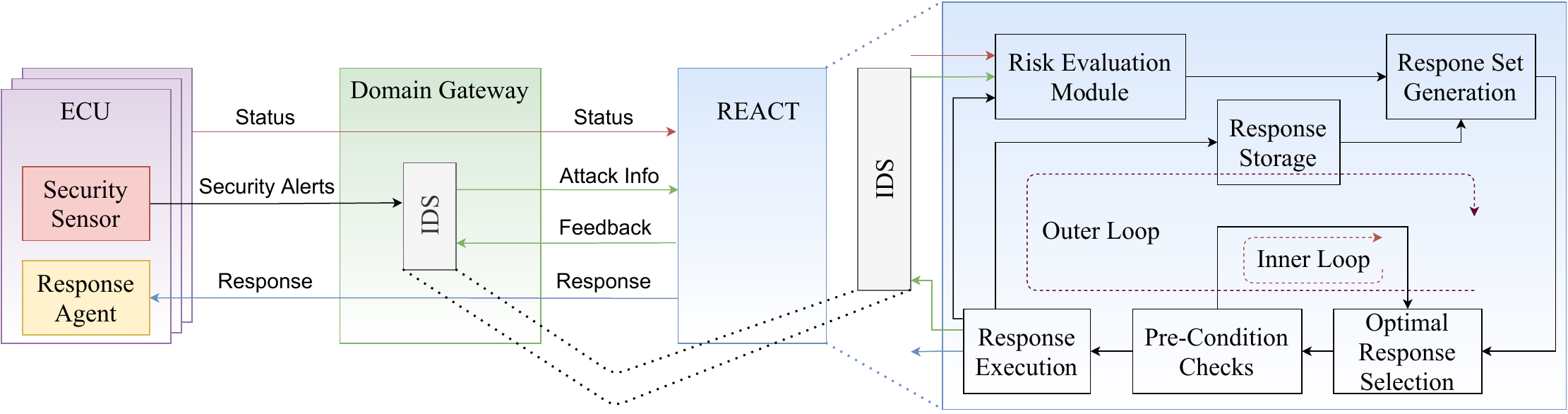}
		\caption{Internal architecture of REACT.}
	\label{fig:int_IRS_architecture}
\end{figure*}

In this section, we will discuss some design decisions regarding  REACT, our proposed automotive \ac{IRS} (refer to \cref{sec_irsdeployment}) and detail its components (refer to \cref{sec:irs_components}). 

\subsection{IRS Deployment}
\label{sec_irsdeployment}
Our proposed automotive \ac{IRS} can be deployed in three different locations: 
\begin{itemize}
	\item Central Gateway: The vehicle will have one \ac{IRS} that receives information from various \acp{ECU}. This central \ac{IRS} will have a comprehensive view and understanding of the entire system. However, it is considered a single point of failure.  
	\item Domain Gateway: The vehicle will have one \ac{IRS} per domain gateway. Each one will be mainly responsible for the \acp{ECU} belonging to that domain and will interact with other \acp{IRS}. Implementing this solution requires the existence of an Intrusion Response eXchange Protocol (IRXP) \cite{1}. 
	\item  \ac{ECU}:   The vehicle will have one \ac{IRS} per \ac{ECU}. This \ac{IRS} will be primarily responsible for reacting to attacks related to its host \ac{ECU}. Simultaneously, it can exchange responses related to other \acp{ECU} if needed. Choosing this option ensures the absence of a single point of failure. However, deploying such a solution requires that each \ac{ECU} is capable of running the \ac{IRS}, and it also necessitates the existence  and the support of an IRXP \cite{1}. 
 
\end{itemize}

The architecture depicted in \figurename~\ref{fig:int_IRS_architecture} illustrates the scenario where the \ac{IRS} is deployed in the central gateway. Any potential change would be primarily associated with the source of certain information required for the functionality of the \ac{IRS}, whether it originates from the same \ac{ECU} (in the case of implementing the IRS per \ac{ECU}) or from external sources such as other \acp{ECU} or domains at the gateway.
Regardless of the chosen deployment location for the \ac{IRS}, it necessitates the reception and sharing of information with other components within the vehicle, as outlined below: 

\begin{itemize}
	\item Attack Information: This information is provided by the \ac{IDS}. In our research, we consider the \ac{IDS} functionality as trusted, treating it as a black-box that reliably detects intrusions without requiring additional false-positive handling \cite{24, 127}. In our architecture, we place the \ac{IDS} in the domain gateway. 
	Consequently, a security sensor \cite{10} is needed to monitor its portion of the environment for security-related observations. This data is then reported to the domain-specific gateway, which houses the domain \ac{IDS}. 
	\item Status Information: This includes information about the various states of the vehicle and its surroundings. This data is collected and aggregated from various vehicle sensors and shared with the \ac{IRS}.
	\item Response Information: This information can encompass the precise responses needed for specific \ac{ECU}s or those that need to be shared with the \ac{SOC}. In our architecture, we assume the presence of response agents located in each \ac{ECU}. These agents are responsible for receiving responses and deploying them within the respective \ac{ECU}.
	
\end{itemize}
It is crucial to mention the necessity of ensuring the security of this data by implementing secure communication between the \ac{ECU}, domain gateway, and the \ac{IRS}. 
\subsection{IRS component}
\label{sec:irs_components}
The \ac{IRS} consists of the following sub-components (as shown in \figurename~\ref{fig:int_IRS_architecture}):

\begin{itemize}
\item Risk Evaluation Module: This module will be responsible for assessing the impact of an intrusion. The component will receive information about the intrusion from the \ac{IDS} as well as information about the vehicle status.  
\item Response Set Generation: This module compiles a list of possible responses, utilizing information obtained from both the \ac{IDS} and the risk evaluation module. Please note that not every response is applicable to every type of intrusion result (refer to Table \ref{tab:responses}).

\item Optimal Response Selection: This component integrates data from all previous modules to determine the optimal response that can be applied. Within this component, any of the algorithms presented in \cref{sec:posiblealgo} can be integrated. 

\item Precondition Checking: Given the limitations imposed by the system architecture, where not all types of responses can be applied (for example, in cases where a sensor is unavailable due to a DoS attack, it may not always be possible to use a redundant source of information from another sensor if such a backup sensor does not exist), it is imperative to verify whether the selected optimal response is applicable or if an alternative response must be chosen. The Precondition Checking module receives the chosen response and assesses its feasibility. If a response is found to be inapplicable, a feedback loop is established with the previous Optimal Selection Module. This \textit{inner loop} is repeated until the necessary preconditions for an individual response are met.
The order of the Optimal Response Selection and the Precondition Checking is carefully evaluated and results in time benefits:
\begin{enumerate}
	\item "Check-First-Then-Select": The logical order of first eliminating all inapplicable responses and subsequently selecting the best response $r$ from the remaining available options is illustrated by the timing behavior of Equation~\ref{eq:check_select}. 
	\begin{equation}
	\label{eq:check_select}
	t=\left (\sum_{i=1}^{|\mathcal{R}|}t_{check,r_i}\right )+t_{select,r}+t_{execute,r}
	\end{equation}
	The time to select the optimal response $t_{select,r}$ and the time to execute the response $t_{execute,r}$ are summed only once, since the selected response will satisfy the preconditions. In contrast, the time to check the preconditions $t_{check,r}$ is summed over the set of possible responses $\mathcal{R}$, since every response's precondition will be checked. 
	\item "Select-First-Then-Check": While a response may be applied with the probability $p$, it might also be that the constraints are not satisfied with a probability $(1-p)$. This leads to a timing behavior of Equation~\ref{eq:select_check}.
	\begin{equation}
	\label{eq:select_check}
	\begin{aligned}
	\begin{multlined}
	t=t_{select,r_1}+t_{check,r_1}+p\cdot t_{execute,r_1}
	+(1-p)\\ \cdot\sum_{i=2}^{|\mathcal{R}|}\left ( t_{select,r_i}+t_{check,r_i} \right )
	\end{multlined}
	\end{aligned}
	\end{equation}
	While the first selected response must always be checked, it is only executed with the probability $p$. If the preconditions are not satisfied, the \textit{Inner Loop} will be repeated maximum $|\mathcal{R}|-1$ times.
\end{enumerate}
It is evident that for a certain number of responses approaching infinity, Equations~\ref{eq:check_select} and~\ref{eq:select_check} yield the same runtime $t$ when $p=0.5$.
For higher values of $p$, the runtime as per Equation~\ref{eq:select_check} is even lower. This holds true even when $t_{select,r}$ decreases, as the number of possible responses decreases accordingly. Based on these equations, the architecture depicted in Figure~\ref{fig:int_IRS_architecture} exhibits a "Select-First-Then-Check" behavior.

\item Response Execution: This component is responsible for transmitting the chosen response initially to the domain-specific gateways and subsequently to the respective \acp{ECU} for implementation through their local response engines. After a predefined duration, this component triggers the \ac{IDS} to assess the effectiveness of the applied response in mitigating the intrusion. By incorporating this \ac{IDS}-Feedback loop, the \textit{Outer Loop} can be iterated multiple times, each iteration involving a system re-evaluation. This concept serves to counter persistent attacks or stepping-stone attacks effectively. Furthermore, the feedback loop can be utilized to update the parameters of the risk evaluation module for addressing future intrusions.

An essential consideration in the \ac{IRS} architecture shown in Figure~\ref{fig:int_IRS_architecture} is the implementation of  termination criteria for the \textit{inner} and \textit{outer loop}. The absence of such criteria could lead to an endless loop, posing a risk to the stability of the entire \ac{IRS} system. While some prior research has addressed termination criteria~\cite{1, 9}, these methods often involve complex evaluation techniques~\cite{42, 46} or rely on artificial intelligence support~\cite{55}. However, the high computational requirements and intricate modeling approaches associated with these methods are impractical for automotive infrastructure. To address the challenge of preventing endless loops in both the \textit{inner} and \textit{outer} loops, we employ two distinct methods.
	\begin{enumerate}
		\item Preventing Inner Endless Loops: To avoid an endless evaluation of preconditions, we continuously reduce the possible response set by eliminating non-applicable responses. Additionally, we have introduced a special response, labeled as "No Action" (indexed as 31), which will consistently lead to the last possible response. This specific response carries the highest cost, similar to the impact of an intrusion, but provides no benefit. These attributes ensure that the \textit{inner loop} never reaches a deadlock since "No Action" can always be applied.
			\item Avoiding Outer Endless Loops: Once a response is applied, the system undergoes an analysis through the \ac{IDS}-Feedback mechanism to identify if a new stepping-stone attack is detected or if the system is secure. In case a new stepping-stone attack is detected, the entire \textit{outer loop} illustrated in \figurename~\ref{fig:int_IRS_architecture} reiterates. To prevent an endless loop scenario when the same response is repeatedly applied, we implement changes to the parameters of the applied response based on the success of the response. The parameter adaptation differs between a successful and a non-successful response. When the selected response is unsuccessful, it indicates that the benefit values assigned to all HEAVENS parameters may not be accurate. Consequently, an adjustment is needed, resulting in a reduction of the benefit values for all HEAVENS parameters in the previously applied response. This entails the assumption that the relative order of each parameter remains unchanged; for example, if the safety benefit held a higher value than the financial benefit prior to the adjustment, it will continue to do so afterward. This behavior is mathematically expressed in Equation \ref{eq:adaptParam_notsuccessful}.
			 
		\begin{equation}
		\begin{aligned}
		\label{eq:adaptParam_notsuccessful}
		\forall i \in \{S, F, O, P\} &:\\ i_{\text{new}}(i_{\text{old}}) = 
		& \begin{cases}
		10, & \text{if } i_{\text{old}} = 100 \\
		1, & \text{if } i_{\text{old}} = 10 \\
		0, & \text{if } i_{\text{old}} = 1 \text{ or } i_{\text{old}} = 0
		\end{cases}
		\end{aligned}
		\end{equation}
A similar parameter adaptation is required in case the response was applied successfully. However, the parameters cannot simply be increased, as this could lead to predictable responses. Predictable responses pose security risks, as attackers can exploit this behavior~\cite{5}. For that reason, two adaptations are made if the response is successful to avoid predictable behavior:
\begin{itemize}
	\item Original values are restored if the response was previously not successful and its values were adapted according to Equation~\ref{eq:adaptParam_notsuccessful}.	
	\item In a second step, the corresponding weights $w_{i\in {S, F, O, P}}$ are randomly adjusted using a prefactor $r$, where $r_{min} \leq r \leq r_{max}$. This retains the original order of magnitude of $w_i$ while introducing sufficient variation through the multiplication $r\cdot w_i$ to generate different results in the next iteration.
\end{itemize}	
Note that Equation \ref{eq:adaptParam_notsuccessful} presented earlier does not account for the dynamic environmental parameter, denoted as $E$, and its corresponding weight, $w_E$. Further details and definitions are necessary to incorporate this parameter into the adaptation process. These details should encompass various aspects of the vehicle's status and its surrounding environment. For simplicity, we have focused on the vehicle's velocity as a parameter that can help represent the vehicle's status.  	To determine a realistic rating for the impact of vehicle speed, several factors must be taken into account. Studies of traffic accidents have revealed that the impact is influenced not only by the types of vehicles involved but also by their positions at the potential crash site~\cite{133}. Additionally, the age of the passengers in the vehicles can affect the impact of injuries in a traffic accident~\cite{134}. Based on this research, the approach presented in Equation~\ref{eq:velocitySeverity} is applied to the parameter $E$ in the adapted HEAVENS method's prototype implementation~\cite{133, 134}.
\begin{equation}
\label{eq:velocitySeverity}
E(v) = 
\begin{cases}
100,& \text{if } v\geq 75 ~{km/h}\\
10,& \text{if } 50 ~{km/h} \leq v < 75  ~{km/h}\\
1,& \text{if } 30 ~{km/h} \leq v < 50  ~{km/h}\\
0,& \text{if } 0 ~{km/h} \leq v < 30  ~{km/h}
\end{cases}
\end{equation}
	\end{enumerate}	

\item Response Storage: Within this component, a repository is maintained containing a range of potential responses alongside their associated metrics. These metrics can be updated through the feedback mechanism or expanded with the inclusion of new responses and parameters via an external connectivity interface. When implementing this on specific hardware, it is crucial to implement security measures to prevent unauthorized tampering with the memory area.
\end{itemize}

Our proposed \ac{IRS} architecture, featuring both an \textit{inner loop} and an \textit{outer loop}, coupled with the incorporation of automotive-specific considerations into the external architecture, introduces a novel paradigm in the realm of fully automated \acp{IRS}. 
Note that there is already some related work for each part of the IRS (such as the selection method), which was covered in the previous sections. However, there is no system that attempts to include all the aspects against which we can compare our work.

%% file: sections/eval.tex
\section{Evaluation}
\label{sec:eval}

\begin{table*}[ht!]
	\caption{IDS-related information and vehicle state parameters for both evaluation scenarios. }
		\label{table:scenarios}
	\begin{tabular*}{0.99\linewidth}{@{\extracolsep{\fill}} p{0.2\linewidth}p{0.3\linewidth}p{0.4\linewidth}@{}}
		\toprule
		\textbf{Property}  & \textbf{Scenario 1} & \textbf{Scenario 2} \\
		\midrule
		\textbf{Name}  & Adversarial sample &  Information disclosure at the infotainment system \\
		
		\textbf{Infected Asset}  & Front Camera & Infotainment Gateway \\
		\textbf{Affected Asset}  & Acceleration control & Infotainment Gateway \\
		
		\textbf{Intrusion Result}  &Falsify / Alter behavior & Information Disclosure \\
		\textbf{Dynamic Parameter}  &Velocity: 70 $km/h$ &  Velocity: 0 $km/h$ \\
		\bottomrule
	\end{tabular*}
\end{table*}

\subsection{Implementation, Testbed, and Use Cases}
The proposed IRS was implemented using the Python programming language. To implement Linear Programming and the associated Simplex algorithm, we utilized the \texttt{PuLP library} \cite{132}, a well-established choice, along with the GNU Linear Programming Kit as the solver. It is important to note that the adapted \ac{SAW} method remains independent of this decision, as it relies solely on standard Python mathematical operators.

The testbed designed for evaluating the Intrusion Response Engine (IRE) incorporates an embedded system setup to realistically emulate the automotive infrastructure. To ensure this fidelity, our implementation was executed on a Raspberry Pi 4 Model B Rev 1.2, a choice justified by the device's ARM-based quad-core processor running at \SI{1.5}{\giga\hertz}. This processing power closely aligns with the high-performance chips commonly found in the automotive industry. 
	
	The goal of the evaluation is to assess two key aspects of the proposed \ac{IRS}. Firstly, we aim to evaluate its proficiency in optimal response selection, and secondly, we intend to measure various computational metrics, including memory consumption and  the  time required to obtain optimal responses while using the three different selection algorithms:  LP with maximum benefit, LP with minimum cost, and adapted \ac{SAW}.

For our evaluation, we employed two representative intrusion scenarios inspired by real-world intrusions:
	\begin{enumerate}
		\item Adversarial Sample: This scenario involves slight modifications to the input data of a machine learning algorithm, resulting in significantly different outputs from the original~\cite{81}. Given the prevalent use of machine learning algorithms in cameras for automated vehicles, they are vulnerable to exploitation via adversarial samples~\cite{81}. In our evaluation, we exploited a front camera in a rural setting, leading to an altered behavior in the acceleration control. 
			\item Information Disclosure at the Infotainment System: This scenario draws inspiration from an actual attack on a vehicle, where an information disclosure in the infotainment system served as the initial step in a stepping-stone attack~\cite{83}.
	\end{enumerate}
The specific \ac{IDS} parameters and vehicle states employed as input for the scenarios are meticulously detailed in Table \ref{table:scenarios}. 
Please remember that in our prototype of the \ac{IRS}, we consider only the velocity of the attacked vehicle as an illustrative example of a vehicle's status.
	
\subsection{Results}
In this section, we will present the results of testing our \ac{IRS} using two prominent scenarios. We will evaluate response quality, response selection time, memory consumption, and the adaptation of response parameters for each of the three selection algorithms: LP with maximum benefit, LP with minimum cost, and the adapted \ac{SAW}.


\subsubsection{Response Quality}
\label{sec:eva_Re_quality}
\input{plots/fig1.tex}
\input{plots/fig2.tex}
The objective of the response quality evaluation is to assess how different optimal selection algorithms prioritize responses and determine the overall impact and benefit of the applied responses. 
To achieve that, the precondition of each response is set to 'rejected' for every proposed response. This ensures that the \ac{IRS} will continue to suggest responses from the list of possible responses.
Each applied response can have both positive and negative effects on the system, so the cost and benefit values of the selected responses are presented. In this evaluation, default parameters are utilized for each new test, ensuring uniformity in the algorithm evaluation across various metrics.  

Figure~\ref{fig:qualityevalallno} depicts the cost and benefit of all proposed responses in the order they are applied by the respective algorithm for both scenarios. 
The figure shows that our proposed \ac{IRS} suggests a different number and order of responses for various scenarios and for different selection algorithms within the same scenario. Please note that the figure shows that some responses were selected twice. For example, the response of restarting the misbehaving system (indexed with number 19, see Table ~\ref{tab:responses}), was selected twice. However, it is important to clarify that the response was selected for different systems. In other words, the first restart is related to the camera, while the second is for the acceleration control. 
In addition, as expected and shown in \figurename~\ref{fig:qualityeval-lpmax-sc1} and \figurename~\ref{fig:qualityeval-lpmax-sc2}, the LP method with maximum benefit starts at very high benefits. Similarly, the LP with minimum response costs starts at a very low cost and more expensive responses are not selected until later stages, as shown in  \figurename~\ref{fig:qualityeval-lpmin-sc1} and \figurename~\ref{fig:qualityeval-lpmin-sc2}. 
Notably, the LP with maximum benefit operates independently of the cost. However, it always ensures that the cost of the response is less than the impact of the intrusion (see Equation ~\ref{eq:adoptLP_constraint1}).

The reason for the arbitrary behavior is that Linear Programming only follows one optimization function and just satisfies the constraints, but does not sort by constraints.  Similarly, LP with minimum cost delivers arbitrary values with respect to the benefit because it only considers cost metrics in its optimization. 
While the LP with the minimum cost provides more conservative solutions, the LP with maximum benefit suggests more offensive solutions. In a real-world scenario, LP with minimum cost might require multiple responses since its benefits are arbitrarily sorted, while LP with maximum benefit might require more iterations of the "inner loop" since the preconditions for more offensive responses might not be fulfilled.

The adapted \ac{SAW} method exhibits a similar arbitrary behavior as shown in \figurename~\ref{fig:qualityeval-saw-sc1} and \figurename~\ref{fig:qualityeval-saw-sc2}. 
However, it is noticeable that adapted \ac{SAW} may select responses with a cost higher than the impact of the intrusion (see \figurename~\ref{fig:qualityeval-saw-sc2}).
Given that the adapted \ac{SAW} method does not consider constraints, it is an unattractive solution to use any \ac{SAW} method in an automatic \ac{IRS}.

\subsubsection{Time of Response Selection} 
 
To evaluate the time required for selecting a response from a given response list using the selection algorithms, we utilized the previously described method where the \textit{inner loop} of the \ac{IRS} repeats multiple times. 
It is important to note that the generation of the response set occurs only once for an individual intrusion. The time required for list generation is independent of the selection algorithm, measuring at \SI{4.32}{\milli\second} for scenario 1 and \SI{3.82}{\milli\second} for scenario 2. The difference in the measured time between the scenarios is due to the variation in number of possible responses.

\figurename~\ref{fig:qualityevaltime} illustrates the time consumed by the three selection algorithms during the process of selecting different responses. Please note that the X-axis represents the order of the response, not the index of the response.
The figure indicates that the adapted \ac{SAW} method consumes less time compared to the LP methods. Specifically, the LP method with maximum benefit typically consumes more time due to the need for multiple iterations, as its offensive responses may not meet necessary preconditions. Slightly less time is needed for the LP method with minimum cost, although its conservative responses are selected after fewer precondition checks. 
Overall, all algorithms demonstrate good performance on a resource-constrained embedded system.

\subsubsection{Memory Consumption} To measure memory consumption, we utilized Python's internal \texttt{resource} module~\cite{138}. Since some of the optimal selection algorithms rely on third-party libraries, the assessment of memory consumption includes the memory allocated for these functionalities as well. The results are presented in Table~\ref{tab:eval_memory}. The results show that both  LP with maximum benefit and LP with minimum cost methods consume nearly the same amount of memory, while the adapted \ac{SAW} method exhibits considerably lower memory consumption. This difference can be attributed to the external libraries \texttt{PuLP} and the \texttt{GNU Linear Programming Kit}, which require more memory due to their complex data structures and solving methods. Nevertheless, all three selection algorithms exhibit low memory consumption, making them suitable for use in resource-constrained embedded hardware systems.

\begin{table}[t!]
	\centering
	\caption{Memory consumption of the \ac{IRS} in \SI{}{\kilo\byte} using static evaluation.}
	\label{tab:eval_memory}
	\begin{tabular}{p{0.2\columnwidth}p{0.2\columnwidth}p{0.2\columnwidth}p{0.2\columnwidth}}\\
		\toprule
		& LP with Max Benefit & LP with Min Cost & Adapted \ac{SAW}\\ 
		\midrule
		Scenario 1 & \si{19308}                  & \si{19206 }               & \si{11296}  \\ 
		Scenario 2 & \si{19228 }                  & \si{19344 }               & \si{11220}  \\ 
		\bottomrule
	\end{tabular}
\end{table}

\subsubsection{Dynamic Evaluation}
 \input{plots/dynamic1.tex}
 \input{plots/dynamic2.tex}
 The dynamic evaluation concentrates on two key aspects: response and threat impact parameters adaptation (refer to \cref{sec:TARA_formalizationofintrusionsandresponses}) and the inclusion of velocity considerations (as shown in Equation ~\ref{eq:velocitySeverity}). When it comes to parameters adaptation, response quality is assessed based on their cost and benefit. In terms of velocity, we evaluate response variation. These assessments are conducted for both scenarios 1 and 2. By testing all three implemented optimal selection algorithms, we can compare their dynamic behavior.
 \paragraph{Parameters adaption}

To assess the impact of changing parameters, we conducted two repetitions of each scenario, each comprising five iterations of the \textit{outer loop}. In one set of iterations for each scenario, we consistently deemed the responses as successful, while in the other set of five iterations, the responses were uniformly considered unsuccessful.
The benefits and costs of the five optimally selected responses for both scenarios, as determined by the three selection algorithms, under the assumption that the responses were always successful, are presented in \figurename~\ref{fig:dynamicqualityevalayes}. Correspondingly, the results under the assumption that the responses were consistently unsuccessful are displayed in \figurename~\ref{fig:dynamicqualityevalano}.

In consistently successful attacks, we observed that parameter weights change within the range of ±20\% (we have selected $r_{min} = 0.8$ and $r_{max} = 1.2$). The purpose of these changes was to reduce response predictability. In both scenarios, changes in response benefit were evident. However, in the first scenario, all three algorithms retained the same response as shown in \figurename~\ref{fig:dynamicqualityeval-lpmax-sc1}, ~\ref{fig:dynamicqualityeval-lpmin-sc1}, and ~\ref{fig:dynamicqualityeval-saw-sc1}. This was changed in the second scenario, where responses were altered for the LP with maximum benefit and adaptive SAW algorithms as shown in \figurename~\ref{fig:dynamicqualityeval-lpmax-sc2}, and ~\ref{fig:dynamciqualityeval-saw-sc2}. The reason for the absence of changes in the selected responses in the first scenario when using LP with maximum benefits or adapted SAW algorithms can be attributed to the specific response chosen: transitioning to a safe mode (indexed with 17). This response had very high benefit values, as determined through the initial evaluation process, making minor variations of ±20\% inconsequential to the overall result.  Consequently, minor variations of ±20\% did not affect the overall result, as the next possible response had significantly lower benefit values. 
To avoid such a constant behavior, a more substantial modification of the response parameters or the use of an asymmetric window for the prefactor, with a higher probability of negative values, can be implemented. 
Notably, the LP method with minimum cost (\figurename~\ref{fig:dynamicqualityeval-lpmin-sc1} and ~\ref{fig:dynamciqualityeval-lpmin-sc2}) did not consider response benefits in its optimization function, rendering modifications to response benefit irrelevant. This method-related limitation persisted across both simulated scenarios.

In the case of consistently unsuccessful attacks, we observe more substantial variations in the selected responses compared to the previous case (see  \figurename~\ref{fig:dynamicqualityevalano}). This behavior is expected, as the parameter adaptation in a non-successful case involves higher orders of magnitude, as shown in Equation~\ref{eq:adaptParam_notsuccessful}, compared to the successful case. 
Similar to the previous analysis, the LP method with minimum cost optimization consistently generates the same response due to the exclusion of response benefit in the optimization process, as shown in Figures~\ref{fig:dynamicqualityeval-lpmin-no-sc1} and~\ref{fig:dynamciqualityeval-lpmin-nosc2}. Conversely, LP with maximum benefit optimization aligns with expectations. Although the initial response is similar to the successful case, subsequent responses exhibit lower benefits (Figures~\ref{fig:dynamicqualityeval-lpmax-no-sc1} and~\ref{fig:dynamicqualityeval-lpmax-no-sc2}) and higher costs as a side effect. Notably, response index 26 (killing the process) appeared twice in Figures~\ref{fig:dynamicqualityeval-lpmax-no-sc1} and~\ref{fig:dynamicqualityeval-saw-no-sc1}, each referring to different components (i.e., camera and acceleration control). The adapted SAW method consistently produces varying results with less distinct trends in benefit and cost when compared to LP with maximum benefit (Figures~\ref{fig:dynamicqualityeval-saw-no-sc1} and~\ref{fig:dynamciqualityeval-saw-no-sc2}). This observed behavior holds true for both scenarios1 and 2, underscoring the expected functionality of parameter adaptation for non-successful cases.

In conclusion, this assessment of dynamic parameter adaptation confirms that LP with maximum benefit and the adapted SAW methods perform effectively with adjusted parameters, rendering the results valid for both test cases. On the other hand, the LP method with minimum cost optimization falls short in its capacity to respond to parameter shifts in response benefit values. Consequently, this method appears less appealing for identifying optimal responses in autonomous IRS.

\paragraph{Inclusion of Velocity Considerations}
  
The second key aspect of dynamic evaluation involves assessing the influence of vehicle velocity on the selected responses. In our current prototype system, the environmental parameter $E$ is treated similarly to other HEAVENS parameters in Equation~\ref{eq:HEAVENS_adapted}, as their respective weights $w$ are either one or zero. As we alter the velocity, the environmental parameter for an intrusion takes on different values, as indicated in Equation~\ref{eq:velocitySeverity}. Therefore, intrusion's impact is more significant at higher velocities. For this test, both scenario one and two are assessed at three velocities: $0$, $50$, and $100$ km/h, using all three implemented algorithms, with each evaluation beginning with the default data-set.
\begin{table}
	\centering
	\caption{Impact of the velocity for the evaluated scenarios, using Equation~\ref{eq:HEAVENS_adapted}.}
	\label{tab:eval_dynamic_velocity_impact}
	\begin{tabular}{@{}lcccc@{}}
		\toprule
		\multirow{2}{*}{\textbf{}} & \multicolumn{3}{c}{\textbf{Impact (unitless)}} \\
		\cmidrule(lr){2-4}
		& \textbf{0 km/h} & \textbf{50 km/h} & \textbf{100 km/h} \\
		\midrule
		\textbf{Scenario 1} & 200 & 210 & 300 \\
		\textbf{Scenario 2} & 120 & 130 & 220 \\
		\bottomrule
	\end{tabular}
\end{table}
 
While the intrusion impact calculation in Table~\ref{tab:eval_dynamic_velocity_impact} functions as expected, each algorithm consistently selects the same response within each scenario, regardless of the velocity. This behavior can be attributed to the high impact values in the two evaluated scenarios. In cases of less severe intrusions or during the early stages of a stepping-stone attack, where the HEAVENS parameters result in lower values, the velocity's impact becomes relatively more substantial, thus leading to varying results. Nonetheless, it's important to emphasize that the proposed \ac{IRS} architecture is adaptable since the individual weights $w$ for HEAVENS parameters can be customized as per Equation~\ref{eq:HEAVENS_adapted}. This customization minimizes the over-representation of static HEAVENS parameters, enabling the velocity to exert a more pronounced influence on the selected response.
 
\subsubsection{Final Remarks}
The evaluation of the developed \ac{IRS} reveals the advantages and drawbacks of each selection method. 
The adapted \ac{SAW} method is limited by its inability to consider constraints. Consequently, it is not feasible to employ this method in a fully automated \ac{IRS}.
On the other hand, LP with minimum cost consistently favors constant responses and is therefore unsuitable for optimal response identification. Despite its successful application in existing research~\cite{39, 43}, the results demonstrate suboptimal behavior for the automotive use case. Nevertheless, it is well-suited for proposing follow-up responses once the primary intrusion has been mitigated. These follow-up responses can enhance security by alerting a \ac{SOC} and providing information to the car manufacturer, ultimately leading to updated software.
In contrast, the LP method with maximum benefit, excels in all metrics evaluated for an automotive \ac{IRS}. Since it offers responses with high benefits from the outset, it is well-suited to respond to the primary intrusion.

%% file: plots/fig1.tex
\begin{figure*}[t!]
	\centering
		\begin{subfigure}{0.5\linewidth}
			\centering
			\begin{tikzpicture}
			
			\begin{axis}[
			height=0.31\textheight,
			xtick=data,
			xticklabels={1, 18, 17, 9, 3, 20, 20, 26, 26, 21, 22, 25, 32, 33, 8, 30, 5, 7, 19, 19, 7, 10, 29, 27, 28, 24, 23, 31},
			xmin=0, xmax = 28, 
			ymin=0, ymax=350,
			grid=both,
			minor tick num=0,
			major grid style={lightgray, dashed},
			minor grid style={lightgray!25},
			xlabel={Response Index},
			ylabel={Cost/Benefit},
			legend cell align={left},
			xticklabel style={rotate=90, anchor=east},
			]
			
			\addplot[blue,mark=x] coordinates {
				(1, 0)
				(2, 100)
				(3, 20)
				(4, 11)
				(5, 2)
				(6, 101)
				(7, 101)
				(8, 200)
				(9, 200)
				(10, 11)
				(11, 11)
				(12, 101)
				(13, 1)
				(14, 1)
				(15, 20)
				(16, 0)
				(17, 20)
				(18, 101)
				(19, 101)
				(20, 101)
				(21, 101)
				(22, 110)
				(23, 20)
				(24, 2)
				(25, 11)
				(26, 11)
				(27, 101)
				(28, 210)
			};
			\addlegendentry{Cost} 
			\addplot[red, mark=*] coordinates {
				(1, 310)
				(2, 310)
				(3, 220)
				(4, 211)
				(5, 121)
				(6, 120)
				(7, 120)
				(8, 120)
				(9, 120)
				(10, 40)
				(11, 40)
				(12, 40)
				(13, 31)
				(14, 31)
				(15, 31)
				(16, 22)
				(17, 22)
				(18, 21)
				(19, 21)
				(20, 21)
				(21, 21)
				(22, 12)
				(23, 4)
				(24, 3)
				(25, 3)
				(26, 2)
				(27, 1)
				(28, 0)};
			\addlegendentry{Benefit} 
			\addplot[green] coordinates {
				(0, 210)
				(28, 210)};
			\addlegendentry{Impact} 
			\end{axis}
			\end{tikzpicture}
			\caption{LP Max Benefit}
			\label{fig:qualityeval-lpmax-sc1}
		\end{subfigure}%
		\hfill
		\begin{subfigure}{0.5\linewidth}
			\centering
			\begin{tikzpicture}
			
			\begin{axis}[
			height=0.31\textheight,
			xtick=data,
			xticklabels={6, 11, 12, 3, 20, 21, 22, 25, 13, 32, 33, 14, 30, 7, 19, 10, 29, 27, 28, 24, 23, 31},
			xmin=0, xmax = 22, 
			ymin=0, ymax=350,
			grid=both,
			minor tick num=0,
			major grid style={lightgray, dashed},
			minor grid style={lightgray!25},
			xlabel={Response Index},
			ylabel={Cost/Benefit},
			legend cell align={left},
			xticklabel style={rotate=90, anchor=east},
			]
			
			\addplot[blue,mark=x] coordinates {
				(1, 20)
				(2, 2)
				(3, 11)
				(4, 2)
				(5, 101)
				(6, 11)
				(7, 11)
				(8, 101)
				(9, 11)
				(10, 1)
				(11, 1)
				(12, 20)
				(13, 0)
				(14, 101)
				(15, 101)
				(16, 110)
				(17, 20)
				(18, 2)
				(19, 11)
				(20, 11)
				(21, 101)
				(22, 120)
			};
			\addlegendentry{Cost} 
			\addplot[red, mark=*] coordinates {
				(1, 121)
				(2, 121)
				(3, 121)
				(4, 120)
				(5, 40)
				(6, 40)
				(7, 40)
				(8, 31)
				(9, 31)
				(10, 31)
				(11, 31)
				(12, 22)
				(13, 21)
				(14, 21)
				(15, 12)
				(16, 4)
				(17, 3)
				(18, 3)
				(19, 2)
				(20, 1)
				(21, 0)	
			};
			\addlegendentry{Benefit} 
			\addplot[green] coordinates {
				(0, 120)
				(28, 120)};
			\addlegendentry{Impact} 
			\end{axis}
			\end{tikzpicture}
			\caption{LP Max Benefit}
			\label{fig:qualityeval-lpmax-sc2}
		\end{subfigure}%
		
		\vspace{1em} 
		
		\begin{subfigure}{0.5\linewidth}
			\centering
			\begin{tikzpicture}
			
			\begin{axis}[
			height=0.31\textheight,
			xtick=data,
			xticklabels={1, 30, 32, 33, 27, 3, 28, 9, 21, 22, 24, 5, 17, 8, 29, 18, 20, 7, 7, 20, 23, 25, 19, 19, 10, 26, 26, 31},
			xmin=0, xmax = 28, 
			ymin=0, ymax=350,
			grid=both,
			minor tick num=0,
			major grid style={lightgray, dashed},
			minor grid style={lightgray!25},
			xlabel={Response Index},
			ylabel={Cost/Benefit},
			legend cell align={left},
			xticklabel style={rotate=90, anchor=east},
			]
			
			\addplot[blue,mark=x] coordinates {
				(1, 0)
				(2, 0) 
				(3, 1) 
				(4, 1) 
				(5, 2) 
				(6, 2) 
				(7, 11) 
				(8, 11) 
				(9, 11) 
				(10, 11) 
				(11, 11) 
				(12, 20) 
				(13, 20) 
				(14, 20) 
				(15, 20) 
				(16, 100) 
				(17, 101) 
				(18, 101) 
				(19, 101) 
				(20, 101) 
				(21, 101) 
				(22, 101) 
				(23, 101) 
				(24, 101) 
				(25, 110) 
				(26, 200) 
				(27, 200) 
				(28, 210)
			};
			\addlegendentry{Cost} 
			\addplot[red, mark=*] coordinates {
				(1, 310)
				(2, 22)
				(3, 31)
				(4, 31)
				(5, 3)
				(6, 121)
				(7, 3)
				(8, 211)
				(9, 40)
				(10, 40)
				(11, 2)
				(12, 22)
				(13, 220)
				(14, 31)
				(15, 4)
				(16, 310)
				(17, 120)
				(18, 21)
				(19, 21)
				(20, 120)
				(21, 1)
				(22, 40)
				(23, 21)
				(24, 21)
				(25, 12)
				(26, 120)
				(27, 120)
				(28, 0)};
			\addlegendentry{Benefit} 
			\addplot[green] coordinates {
				(0, 210)
				(28, 210)};
			\addlegendentry{Impact} 
			\end{axis}
			\end{tikzpicture}
			\caption{LP Min Cost}
			\label{fig:qualityeval-lpmin-sc1}
		\end{subfigure}%
		\hfill
		\begin{subfigure}{0.5\linewidth}
			\centering
			\begin{tikzpicture}
			\begin{axis}[
			height=0.31\textheight,
			xtick=data,
			xticklabels={32, 33, 27, 11, 3, 28, 12, 13, 21, 22, 24, 14, 6, 29, 7, 20, 23, 25, 19, 10, 31},
			xmin=0, xmax = 21, 
			ymin=0, ymax=350,
			grid=both,
			minor tick num=0,
			major grid style={lightgray, dashed},
			minor grid style={lightgray!25},
			xlabel={Response Index},
			ylabel={Cost/Benefit},
			legend cell align={left},
			xticklabel style={rotate=90, anchor=east},
			]
			
			\addplot[blue,mark=x] coordinates {
				(1, 1)
				(2, 1)
				(3, 2)
				(4, 2)
				(5, 2)
				(6, 11)
				(7, 11)
				(8, 11)
				(9, 11)
				(10, 11)
				(11, 11)
				(12, 20)
				(13, 20)
				(14, 20)
				(15, 101)
				(16, 101)
				(17, 101)
				(18, 101)
				(19, 101)
				(20, 110)
				(21, 120)
			};
			\addlegendentry{Cost} 
			\addplot[red, mark=*] coordinates {
				(1, 31)
				(2, 31)
				(3, 3)
				(4, 121)
				(5, 121)
				(6, 3)
				(7, 121)
				(8, 31)
				(9, 40)
				(10, 40)
				(11, 2)
				(12, 31)
				(13, 211)
				(14, 4)
				(15, 21)
				(16, 120)
				(17, 1)
				(18, 40)
				(19, 21)
				(20, 12)
				(21, 0)
			};
			\addlegendentry{Benefit} 
			\addplot[green] coordinates {
				(0, 120)
				(28, 120)};
			\addlegendentry{Impact} 
			\end{axis}
			
			\end{tikzpicture}
			\caption{LP Min Cost}
			\label{fig:qualityeval-lpmin-sc2}
		\end{subfigure}%

		\vspace{1em} 
		
		\begin{subfigure}{0.5\linewidth}
			\centering
			\begin{tikzpicture}
			
			\begin{axis}[
			height=0.31\textheight,
			xtick=data,
			xticklabels={17, 30, 9, 20, 26, 26, 31, 33, 32, 3, 22, 21, 25, 8, 5, 7, 7, 19, 19, 29, 10, 27, 28, 24, 23, 20},
			xmin=0, xmax = 28, 
			ymin=0, ymax=350,
			grid=both,
			minor tick num=0,
			major grid style={lightgray, dashed},
			minor grid style={lightgray!25},
			xlabel={Response Index},
			ylabel={Cost/Benefit},
			legend cell align={left},
			xticklabel style={rotate=90, anchor=east},
			]
			
			\addplot[blue,mark=x] coordinates {
				(1, 20)
				(2, 0)
				(3, 11)
				(4, 101)
				(5, 200)
				(6, 200)
				(7, 210)
				(8, 1)
				(9, 1)
				(10, 2)
				(11, 11)
				(12, 11)
				(13, 101)
				(14, 20)
				(15, 20)
				(16, 101)
				(17, 101)
				(18, 101)
				(19, 101)
				(20, 20)
				(21, 110)
				(22, 2)
				(23, 11)
				(24, 11)
				(25, 101)
				(26, 101)
			};
			\addlegendentry{Cost} 
			\addplot[red, mark=*] coordinates {
				(1, 220)
				(2, 22)
				(3, 211)
				(4, 120)
				(5, 120)
				(6, 120)
				(7, 0)
				(8, 31)
				(9, 31)
				(10, 121)
				(11, 40)
				(12, 40)
				(13, 40)
				(14, 31)
				(15, 22)
				(16, 21)
				(17, 21)
				(18, 21)
				(19, 21)
				(20, 4)
				(21, 12)
				(22, 3)
				(23, 3)
				(24, 2)
				(25, 1)
				(26, 120)};
			\addlegendentry{Benefit} 
			\addplot[green] coordinates {
				(0, 210)
				(28, 210)};
			\addlegendentry{Impact} 
			\end{axis}
			\end{tikzpicture}
			\caption{SAW}
			\label{fig:qualityeval-saw-sc1}
		\end{subfigure}%
		\hfill
		\begin{subfigure}{0.5\linewidth}
			\centering
			\begin{tikzpicture}
			\begin{axis}[
			height=0.31\textheight,
			xtick=data,
			xticklabels={6, 30, 33, 32, 3, 11, 22, 21, 12, 25, 20, 26, 13, 14, 31, 7, 19, 10, 29, 27, 28, 24},
			xmin=0, xmax = 22, 
			ymin=0, ymax=350,
			grid=both,
			minor tick num=0,
			major grid style={lightgray, dashed},
			minor grid style={lightgray!25},
			xlabel={Response Index},
			ylabel={Cost/Benefit},
			legend cell align={left},
			xticklabel style={rotate=90, anchor=east},
			]
			
			\addplot[blue,mark=x] coordinates {
				(1, 20)
				(2, 0)
				(3, 1)
				(4, 1)
				(5, 2)
				(6, 2)
				(7, 11)
				(8, 11)
				(9, 11)
				(10, 101)
				(11, 101)
				(12, 200)
				(13, 11)
				(14, 20)
				(15, 120)
				(16, 101)
				(17, 101)
				(18, 110)
				(19, 20)
				(20, 2)
				(21, 11)
				(22, 11)
			};
			\addlegendentry{Cost} 
			\addplot[red, mark=*] coordinates {
				
				(1, 211)
				(2, 22)
				(3, 31)
				(4, 31)
				(5, 121)
				(6, 121)
				(7, 40)
				(8, 40)
				(9, 121)
				(10, 40)
				(11, 120)
				(12, 120)
				(13, 31)
				(14, 31)
				(15, 0)
				(16, 21)
				(17, 21)
				(18, 12)
				(19, 4)
				(20, 3)
				(21, 3)
				(22, 2)
			};
			\addlegendentry{Benefit} 
			\addplot[green] coordinates {
				(0, 120)
				(28, 120)};
			\addlegendentry{Impact} 
			
			\end{axis}
			\end{tikzpicture}
			\caption{SAW}
			\label{fig:qualityeval-saw-sc2}
		\end{subfigure}
	\caption{Evaluation of the response benefit and cost for Scenario 1 (left) and Scenario 2 (right) using LP with maximum benefit (top), LP with minimum cost (middle), and adapted \ac{SAW} (bottom). }
	\label{fig:qualityevalallno}
\end{figure*}
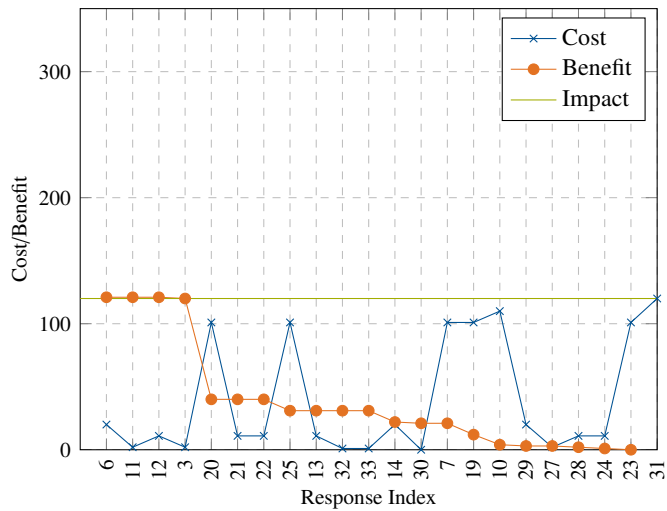
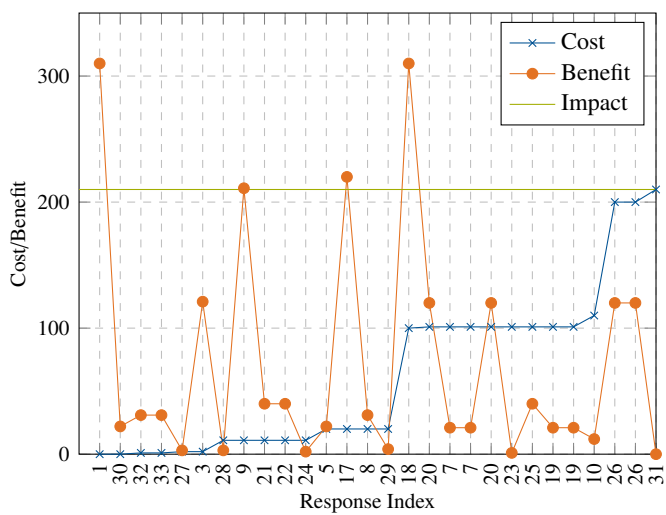
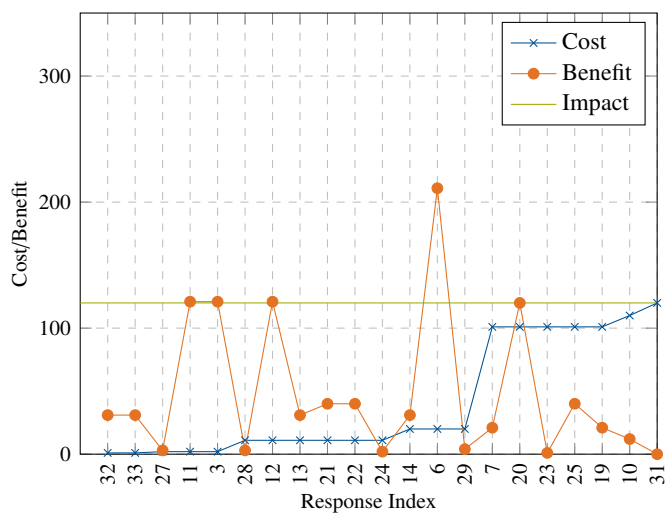
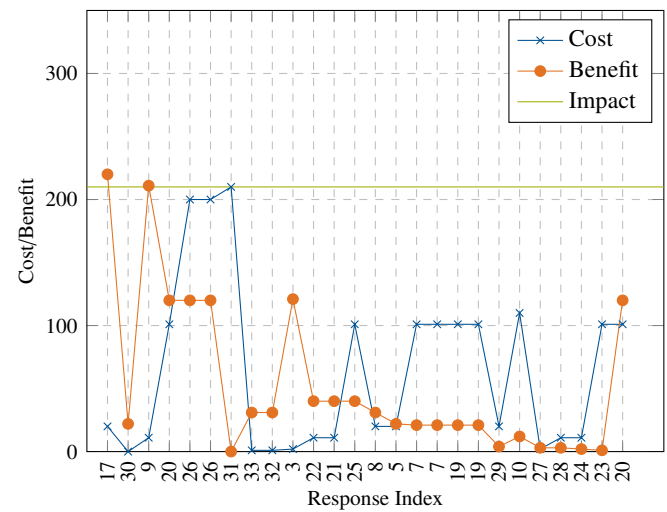
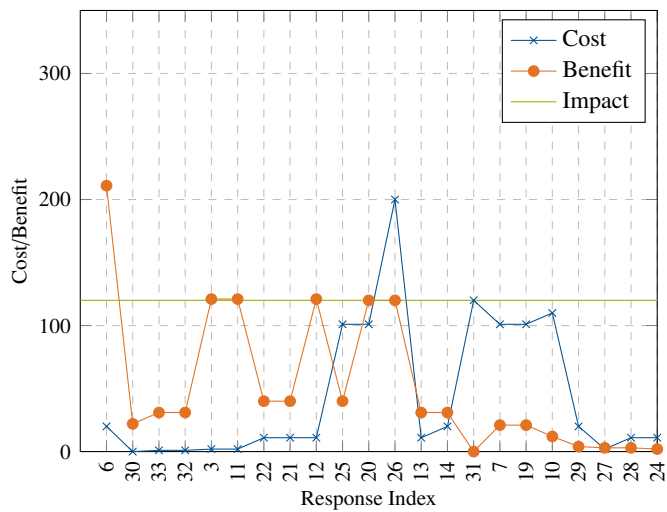

%% file: plots/fig2.tex
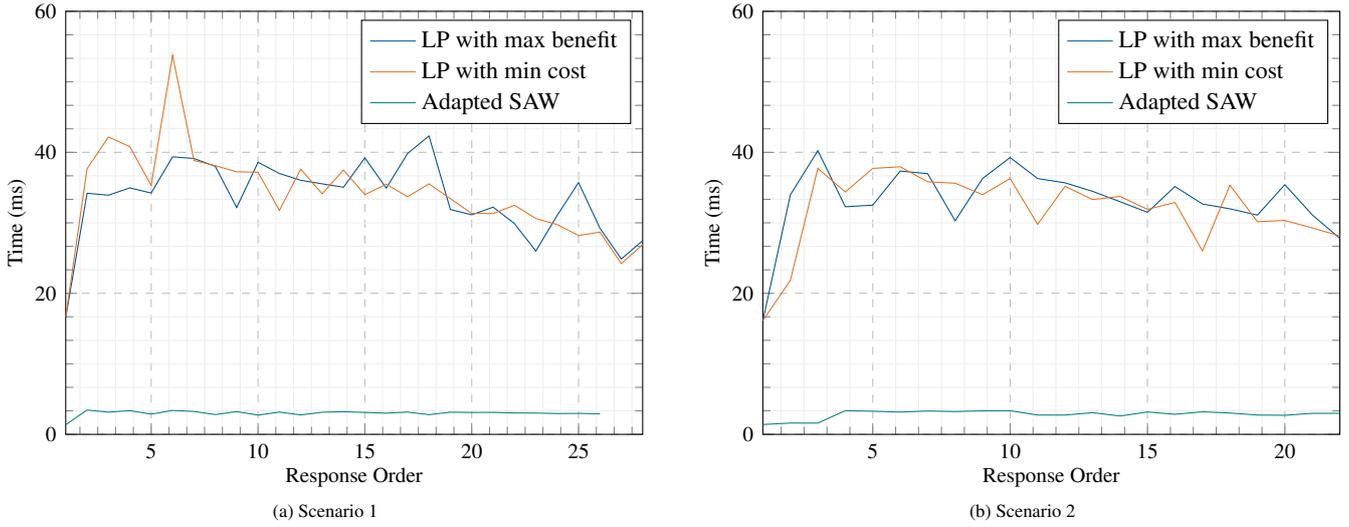
\begin{figure*}[ht!]
	\centering
	
	\begin{subfigure}{0.5\linewidth}
		\centering
		\begin{tikzpicture}
	\pgfplotstableread{plots/static-eval-scenario1-time.dat}\datatableB
	\begin{axis}[
	height=0.3\textheight,
xmin = 1, xmax = 28,
ymin = 0, ymax = 60,
	grid=both,
	minor tick num=5,
	major grid style={lightgray, dashed},
	minor grid style={lightgray!25},
	xlabel={Response Order},
	ylabel={Time (\SI{}{\milli\second})},
	legend cell align={left},
	]
	
	\addplot[blue] table [x = {Response_number}, y = {LP_max_benefit}] {\datatableB};
	\addplot[red] table [x = {Response_number}, y = {LP_min_cost}] {\datatableB};
	\addplot[teal] table [x = {Response_number}, y = {SAW}] {\datatableB};
	
	\legend{
		LP with max benefit, 
		LP with min  cost ,
		Adapted \ac{SAW}
	}
	\end{axis}
	\end{tikzpicture}
		\caption{Scenario 1}
		\label{fig:qualityeval-lpmax-sc1-time}
	\end{subfigure}%
	\hfil
		\begin{subfigure}{0.5\linewidth}
		\centering
		\begin{tikzpicture}
		\pgfplotstableread{plots/static-eval-scenario2-time.dat}\datatableC
		\begin{axis}[
		height=0.3\textheight,
		xmin = 1, xmax = 22,
		ymin = 0, ymax = 60,
		grid=both,
		minor tick num=5,
		major grid style={lightgray, dashed},
		minor grid style={lightgray!25},
		xlabel={Response Order},
		ylabel={Time (\SI{}{\milli\second})},
		legend cell align={left},
		]
		
		\addplot[blue] table [x = {Response_number}, y = {LP_max_benefit}] {\datatableC};
		\addplot[red] table [x = {Response_number}, y = {LP_min_cost}] {\datatableC};
		\addplot[teal] table [x = {Response_number}, y = {SAW}] {\datatableC};
		\legend{
		LP with max benefit, 
		LP with min  cost ,
		Adapted \ac{SAW}
	}
		\end{axis}
		\end{tikzpicture}
		\caption{Scenario 2}
		\label{fig:qualityeval-lpmax-sc2-time}
	\end{subfigure}%
	\caption{Evaluation of consumed time for response selection using the three selection algorithms for both scenarios.}
	\label{fig:qualityevaltime}
\end{figure*}

%% file: plots/dynamic1.tex
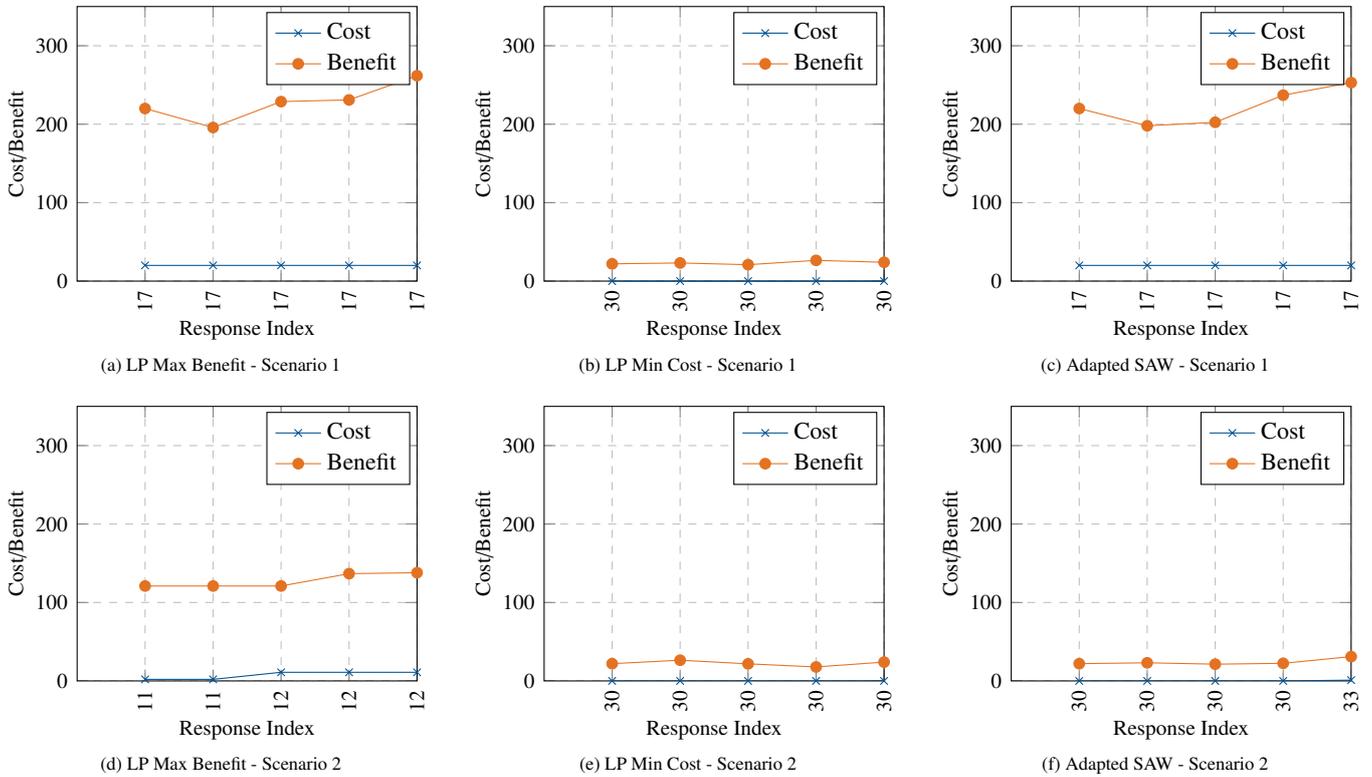
\begin{figure*}[ht!]
	\centering
	
	\begin{subfigure}{0.33\linewidth}
		\centering
		\begin{tikzpicture}
	
	\begin{axis}[
	xtick=data,
	xticklabels={17, 17, 17, 17, 17},
	xmin=0, xmax = 5, 
	ymin=0, ymax=350,
	grid=both,
	minor tick num=0,
	major grid style={lightgray, dashed},
	minor grid style={lightgray!25},
	xlabel={Response Index},
	ylabel={Cost/Benefit},
	legend cell align={left},
	xticklabel style={rotate=90, anchor=east},
	]
	
\addplot[blue,mark=x] coordinates {
	(1, 20)
	(2, 20)
	(3, 20)
	(4, 20)
	(5, 20)
	};
\addlegendentry{Cost} 
\addplot[red, mark=*] coordinates {
  (1, 220)
(2, 195.8)
(3, 228.8)
(4, 231)
(5, 261.8)
};
	\addlegendentry{Benefit} 
	\end{axis}
		\end{tikzpicture}
		\caption{LP Max Benefit - Scenario 1}
		\label{fig:dynamicqualityeval-lpmax-sc1}
	\end{subfigure}%
\hfill
	\begin{subfigure}{0.33\linewidth}
	\centering
	\begin{tikzpicture}
	
	\begin{axis}[
	xtick=data,
  xticklabels={30, 30, 30, 30, 30},
	xmin=0, xmax = 5, 
	ymin=0, ymax=350,
	grid=both,
	minor tick num=0,
	major grid style={lightgray, dashed},
	minor grid style={lightgray!25},
	xlabel={Response Index},
	ylabel={Cost/Benefit},
	legend cell align={left},
	xticklabel style={rotate=90, anchor=east},
	]
	
	\addplot[blue,mark=x] coordinates {
(1, 0)
(2, 0)
(3, 0)
(4, 0)
(5, 0)	};
	\addlegendentry{Cost} 
	\addplot[red, mark=*] coordinates {
	(1, 22)
	(2, 23.1)
	(3, 20.9)
	(4, 26.4)
	(5, 23.98)
	};
	\addlegendentry{Benefit} 

	\end{axis}
	\end{tikzpicture}
			\caption{LP Min Cost - Scenario 1}
	\label{fig:dynamicqualityeval-lpmin-sc1}
	\end{subfigure}%
	\hfill
	\begin{subfigure}{0.33\linewidth}
		\centering
		\begin{tikzpicture}
	
	\begin{axis}[
	xtick=data,
	xticklabels={17, 17, 17, 17, 17},
	xmin=0, xmax = 5, 
	ymin=0, ymax=350,
	grid=both,
	minor tick num=0,
	major grid style={lightgray, dashed},
	minor grid style={lightgray!25},
	xlabel={Response Index},
	ylabel={Cost/Benefit},
	legend cell align={left},
	xticklabel style={rotate=90, anchor=east},
	]
	
	\addplot[blue,mark=x] coordinates {
		(1, 20)
		(2, 20) 
		(3, 20) 
		(4, 20) 
		(5, 20)};
	\addlegendentry{Cost} 
	\addplot[red, mark=*] coordinates {
(1, 220)
(2, 198)
(3, 202.4)
(4, 237)
(5, 253)
};
	\addlegendentry{Benefit} 
	\end{axis}
		\end{tikzpicture}
		\caption{Adapted SAW - Scenario 1}
		\label{fig:dynamicqualityeval-saw-sc1}
	\end{subfigure}%
	\vspace{1em} 
	\begin{subfigure}{0.33\linewidth}
	\centering
	\begin{tikzpicture}
	
	\begin{axis}[
	xtick=data,
	xticklabels={11, 11, 12, 12, 12},
	xmin=0, xmax = 5, 
	ymin=0, ymax=350,
	grid=both,
	minor tick num=0,
	major grid style={lightgray, dashed},
	minor grid style={lightgray!25},
	xlabel={Response Index},
	ylabel={Cost/Benefit},
	legend cell align={left},
	xticklabel style={rotate=90, anchor=east},
	]
	
	\addplot[blue,mark=x] coordinates {
		(1, 2)
		(2, 2)
		(3, 11)
		(4, 11)
		(5, 11)
	};
	\addlegendentry{Cost} 
	\addplot[red, mark=*] coordinates {
		(1, 121)
		(2, 121)
		(3, 121)
		(4, 136.73)
		(5, 137.94)
		};
	\addlegendentry{Benefit} 
		\end{axis}
	\end{tikzpicture}
	\caption{LP Max Benefit - Scenario 2}
	\label{fig:dynamicqualityeval-lpmax-sc2}
\end{subfigure}%
\hfill
\begin{subfigure}{0.33\linewidth}
	\centering
	\begin{tikzpicture}
	
	\begin{axis}[
	xtick=data,
	xticklabels={30, 30, 30, 30, 30},
	xmin=0, xmax = 5, 
	ymin=0, ymax=350,
	grid=both,
	minor tick num=0,
	major grid style={lightgray, dashed},
	minor grid style={lightgray!25},
	xlabel={Response Index},
	ylabel={Cost/Benefit},
	legend cell align={left},
	xticklabel style={rotate=90, anchor=east},
	]
	
	\addplot[blue,mark=x] coordinates {
		(1, 0)
		(2, 0)
		(3, 0)
		(4, 0)
		(5, 0)	};
	\addlegendentry{Cost} 
	\addplot[red, mark=*] coordinates {
		(1, 22)
		(2, 	26.4)
		(3, 21.78)
		(4, 17.82)
		(5, 	23.98)
	};
	\addlegendentry{Benefit} 
	\end{axis}
	\end{tikzpicture}
	\caption{LP Min Cost - Scenario 2}
	\label{fig:dynamciqualityeval-lpmin-sc2}
\end{subfigure}%
\hfill
\begin{subfigure}{0.33\linewidth}
	\centering
	\begin{tikzpicture}
	
	\begin{axis}[
	xtick=data,
	xticklabels={30, 30, 30, 30, 33},
	xmin=0, xmax = 5, 
	ymin=0, ymax=350,
	grid=both,
	minor tick num=0,
	major grid style={lightgray, dashed},
	minor grid style={lightgray!25},
	xlabel={Response Index},
	ylabel={Cost/Benefit},
	legend cell align={left},
	xticklabel style={rotate=90, anchor=east},
	]
	
	\addplot[blue,mark=x] coordinates {
		(1, 0)
		(2, 0) 
		(3, 0) 
		(4, 0) 
		(5, 1)};
	\addlegendentry{Cost} 
	\addplot[red, mark=*] coordinates {
		(1, 22)
		(2, 23.1)
		(3, 21.34)
		(4, 22.44)
		(5, 31)
		};
	\addlegendentry{Benefit} 
	\end{axis}
	\end{tikzpicture}
	\caption{Adapted SAW - Scenario 2}
	\label{fig:dynamciqualityeval-saw-sc2}
\end{subfigure}%
	\caption{ Evaluation of parameter adaptation in Scenario 1 (top) and Scenario 2 (bottom) for the responses selected over five iterations using the three selection algorithms, assuming the responses were consistently considered successful. }
	\label{fig:dynamicqualityevalayes}
\end{figure*}

%% file: plots/dynamic2.tex
\begin{figure*}[th!]
	\centering
	
	\begin{subfigure}{0.33\linewidth}
		\centering
		\begin{tikzpicture}
		
		\begin{axis}[
		xtick=data,
		xticklabels={17, 20, 20, 26, 26},
		xmin=0, xmax = 5, 
		ymin=0, ymax=350,
		grid=both,
		minor tick num=0,
		major grid style={lightgray, dashed},
		minor grid style={lightgray!25},
		xlabel={Response Index},
		ylabel={Cost/Benefit},
		legend cell align={left},
		xticklabel style={rotate=90, anchor=east},
		]
		
		\addplot[blue,mark=x] coordinates {
			(1, 20)
			(2, 101)
			(3, 101)
			(4, 200)
			(5, 200)
		};
		\addlegendentry{Cost} 
		\addplot[red, mark=*] coordinates {
			(1, 220)
			(2, 	120)
			(3, 	120)
			(4, 	120)
			(5, 	120)
				};
		\addlegendentry{Benefit} 
		\end{axis}
		\end{tikzpicture}
		\caption{LP Max Benefit - Scenario 1}
		\label{fig:dynamicqualityeval-lpmax-no-sc1}
	\end{subfigure}%
	\hfill
	\begin{subfigure}{0.33\linewidth}
		\centering
		\begin{tikzpicture}
		
		\begin{axis}[
		xtick=data,
		xticklabels={30, 30, 30, 30, 30},
		xmin=0, xmax = 5, 
		ymin=0, ymax=350,
		grid=both,
		minor tick num=0,
		major grid style={lightgray, dashed},
		minor grid style={lightgray!25},
		xlabel={Response Index},
		ylabel={Cost/Benefit},
		legend cell align={left},
		xticklabel style={rotate=90, anchor=east},
		]
		
		\addplot[blue,mark=x] coordinates {
			(1, 0)
			(2, 0)
			(3, 0)
			(4, 0)
			(5, 0)	};
		\addlegendentry{Cost} 
		\addplot[red, mark=*] coordinates {
			(1, 22)
			(2, 2)
			(3, 0)
			(4, 0)
			(5, 0)
		};
		\addlegendentry{Benefit} 
		
		\end{axis}
		\end{tikzpicture}
		\caption{LP Min Cost - Scenario 1}
		\label{fig:dynamicqualityeval-lpmin-no-sc1}
	\end{subfigure}%
	\hfill
	\begin{subfigure}{0.33\linewidth}
		\centering
		\begin{tikzpicture}
		
		\begin{axis}[
		xtick=data,
		xticklabels={17, 30, 20, 26, 26},
		xmin=0, xmax = 5, 
		ymin=0, ymax=350,
		grid=both,
		minor tick num=0,
		major grid style={lightgray, dashed},
		minor grid style={lightgray!25},
		xlabel={Response Index},
		ylabel={Cost/Benefit},
		legend cell align={left},
		xticklabel style={rotate=90, anchor=east},
		]
		
		\addplot[blue,mark=x] coordinates {
			(1, 20)
			(2, 0) 
			(3, 101) 
			(4, 200) 
			(5, 200)};	
		\addlegendentry{Cost} 
		\addplot[red, mark=*] coordinates {
			(1, 220)
			(2, 22)
			(3, 120)
			(4, 120)
			(5, 120)			
			};
		\addlegendentry{Benefit} 
		\end{axis}
		\end{tikzpicture}
		\caption{Adapted SAW - Scenario 1}
		\label{fig:dynamicqualityeval-saw-no-sc1}
	\end{subfigure}%
	\vspace{1em} 
	\begin{subfigure}{0.33\linewidth}
		\centering
		\begin{tikzpicture}
		
		\begin{axis}[
		xtick=data,
		xticklabels={11, 12, 3, 20, 21},
		xmin=0, xmax = 5, 
		ymin=0, ymax=350,
		grid=both,
		minor tick num=0,
		major grid style={lightgray, dashed},
		minor grid style={lightgray!25},
		xlabel={Response Index},
		ylabel={Cost/Benefit},
		legend cell align={left},
		xticklabel style={rotate=90, anchor=east},
		]
		
		\addplot[blue,mark=x] coordinates {
			(1, 2)
			(2, 11)
			(3, 2)
			(4, 101)
			(5, 11)
		};
		\addlegendentry{Cost} 
		\addplot[red, mark=*] coordinates {
			(1, 121)
			(2, 121)
			(3, 121)
			(4, 120)
			(5, 40)
		};
		\addlegendentry{Benefit} 
		\end{axis}
		\end{tikzpicture}
		\caption{LP Max Benefit - Scenario 2}
		\label{fig:dynamicqualityeval-lpmax-no-sc2}
	\end{subfigure}%
	\hfill
	\begin{subfigure}{0.33\linewidth}
		\centering
		\begin{tikzpicture}
		
		\begin{axis}[
		xtick=data,
		xticklabels={30, 30, 30, 30, 30},
		xmin=0, xmax = 5, 
		ymin=0, ymax=350,
		grid=both,
		minor tick num=0,
		major grid style={lightgray, dashed},
		minor grid style={lightgray!25},
		xlabel={Response Index},
		ylabel={Cost/Benefit},
		legend cell align={left},
		xticklabel style={rotate=90, anchor=east},
		]
		
		\addplot[blue,mark=x] coordinates {
			(1, 0)
			(2, 0)
			(3, 0)
			(4, 0)
			(5, 0)	};
		\addlegendentry{Cost} 
		\addplot[red, mark=*] coordinates {
			(1, 22)
			(2, 	2)
			(3, 0)
			(4, 0)
			(5, 	0)
		};
		\addlegendentry{Benefit} 
		\end{axis}
		\end{tikzpicture}
		\caption{LP Min Cost - Scenario 2}
		\label{fig:dynamciqualityeval-lpmin-nosc2}
	\end{subfigure}%
	\hfill
	\begin{subfigure}{0.33\linewidth}
		\centering
		\begin{tikzpicture}
		
		\begin{axis}[
		xtick=data,
		xticklabels={30, 33, 32, 3, 11},
		xmin=0, xmax = 5, 
		ymin=0, ymax=350,
		grid=both,
		minor tick num=0,
		major grid style={lightgray, dashed},
		minor grid style={lightgray!25},
		xlabel={Response Index},
		ylabel={Cost/Benefit},
		legend cell align={left},
		xticklabel style={rotate=90, anchor=east},
		]
		
		\addplot[blue,mark=x] coordinates {
			(1, 0)
			(2, 1) 
			(3, 1) 
			(4, 2) 
			(5, 2)};
		\addlegendentry{Cost} 
		\addplot[red, mark=*] coordinates {
			(1, 22)
			(2, 31)
			(3, 31)
			(4, 121)
			(5, 121)
			};
		\addlegendentry{Benefit} 
		\end{axis}
		\end{tikzpicture}
		\caption{Adapted SAW - Scenario 2}
		\label{fig:dynamciqualityeval-saw-no-sc2}
	\end{subfigure}%
	\caption{Evaluation of parameter adaptation in Scenario 1 (top) and Scenario 2 (bottom) for the responses selected over five iterations using the three selection algorithms, assuming the responses were consistently considered unsuccessful. }
	\label{fig:dynamicqualityevalano}
\end{figure*}
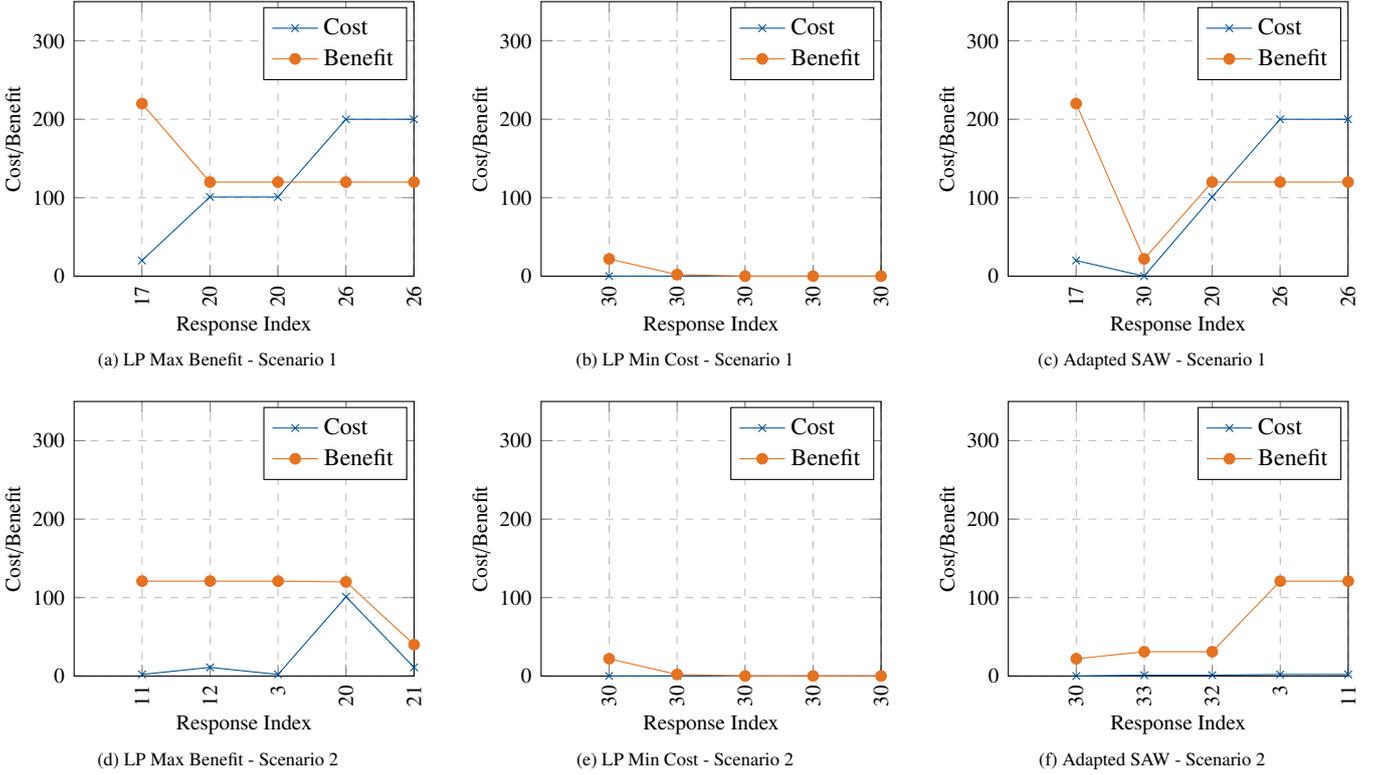

%% file: sections/conclusion.tex
\section{Conclusion and Outlook}
\label{sec:con}
Modern vehicles' intricate architecture and advanced connectivity present unique intrusion challenges. While automotive security research has traditionally emphasized \acp{IDS} as a secondary defense layer, the development of vehicle \ac{IRS} is in its early stages, drawing inspiration from related industries.
To delve into the development of an automotive \ac{IRS}, we sought answers to three key questions: defining potential responses, outlining response evaluation criteria, and optimizing response selection. Initially, we categorized automotive intrusions and stepping-stone attacks into five distinct categories to create a more versatile intrusion model. Similarly, we classified responses, creating a formal description for both intrusions and responses. Additionally, we investigated necessary adjustments to existing risk assessment models to support response evaluation.
Furthermore, we conducted a comprehensive comparison of various optimal selection algorithms, highlighting the adaptability of the \ac{SAW} method and Linear Programming (LP) with various optimizations for \ac{IRS} integration. Although other algorithm families may gain relevance in the future, they currently face limitations in the automotive context.
In addition to these findings, we proposed an \ac{IRS} architecture that accommodates the distributed nature of vehicles and addresses automotive-specific constraints. Evaluation in real-world scenarios has led to the development of a novel vehicular \ac{IRS}, demonstrating its potential for integration into modern distributed vehicle architectures and enhancing overall security.

While the focus of the paper is on the analysis and design of the \ac{IRS}, the implementation of the external architecture and the response execution modules on the local engines on each \ac{ECU} is still a challenge towards an \ac{IRS} as a system.
To test such an overall \ac{IRS} system, real-world data-sets including both normal operation and the attack scenarios are needed. Extensive evaluation in Software-in-the-Loop or Hardware-in-the-Loop testbeds can extend the existing evaluations of algorithms and the overall system.
With respect to the secure communication of intrusions and responses, further research and standardization needs to be performed in order to ensure that the developed \ac{IRS} does not only reply in an adequate manner, but also distributes its responses.
The modular architecture of REACT allows an easy extension towards more complex vehicle architectures and new intrusions or responses. Additionally it allows the integration of new selection algorithms in the future to adapt to possible changed needs.